\documentclass[a4paper,11pt]{article}
\usepackage{jheppub}
\usepackage{mathtools}
\usepackage{amsmath}
\usepackage{bm}
\usepackage{url}
\usepackage{color}
\usepackage{comment}

\newcommand{\LQ}{\Lambda_{\rm QCD}}
\newcommand{\alfs}{\alpha_{s}}
\newcommand{\msbar}{\overline{\rm MS}}

\newcommand{\LMS}{\Lambda_{\rm \overline{MS}}}

\newcommand{\bea}{\begin{eqnarray}}
\newcommand{\eea}{\end{eqnarray}}
\newcommand{\simgt}{\hbox{ \raise3pt\hbox to 0pt{$>$}\raise-3pt\hbox{$\sim$} }}
\newcommand{\simlt}{\hbox{ \raise3pt\hbox to 0pt{$<$}\raise-3pt\hbox{$\sim$} }}

\newcommand{\be}{\begin{equation}}
\newcommand{\ee}{\end{equation}}

\newcommand{\lt}{\left}
\newcommand{\rt}{\right}

\newcommand{\non}{\nonumber \\}

\title{\boldmath\Large
Renormalon Subtraction in OPE by Dual Space Approach:\\
Nonlinear Sigma Model and QCD
}
\author[a]{Yuuki Hayashi,\!}
\author[a]{Go Mishima,\!}
\author[a]{Yukinari Sumino,\!}
\author[b]{HiromasaTakaura}

\affiliation[a]{Department of Physics, Tohoku University,\\Sendai, 980-8578 Japan}
\affiliation[b]{Institute of Particle and Nuclear studies, KEK,\\Tsukuba, 305-0801, Japan}

\emailAdd{yuuki.hayashi.s3@dc.tohoku.ac.jp}

\emailAdd{go.mishima@icloud.com}

\emailAdd{yukinari.sumino.a4@tohoku.ac.jp}

\emailAdd{htakaura@post.kek.jp}

\abstract{
It is becoming 
more important to
subtract renormalons efficiently from perturbative calculations,
in order to achieve high precision QCD calculations.
We propose a new framework ``Dual Space Approach" for renormalon separation,
which enables subtraction of multiple renormalons simultaneously.
Using a dual transform which suppresses infrared renormalons,
we derive a one-parameter integral representation 
of a general observable.
We investigate systematically 
how renormalons emerge and
get canceled in the entire operator product expansion (OPE) of an observable, 
by applying the expansion-by-regions 
(EBR) method 
to this one-parameter integral expression.
In particular we investigate in detail OPEs in
a solvable model, the 2-dimensional $O(N)$ nonlinear $\sigma$ model, by
the dual space approach.
A nontrivial mechanism of renormalon cancellation in this model can be
understood from
an integration identity on which the EBR method is founded.
We demonstrate that the dual space approach can be useful
by a simulation study imitating the QCD case.
Application of
this method to QCD calculations is also discussed.
}

\begin{document}

    \begin{flushright}
      \normalsize TU--1182\\
      \normalsize KEK--TH--2501\\
      \today
    \end{flushright}

\maketitle
\flushbottom
\section{Introduction}
\label{sec1}

After the discovery of the Higgs boson, 
particle physics entered the era of the high-precision LHC
experiments and the Belle II experiment, in which 
it is indispensable to improve
accuracy of the theoretical prediction of QCD, 
to meet demands for the precision frontier physics. 
Due to the asymptotic freedom of QCD, theoretical predictions 
 for high-scale observables can be made more precise by
higher-order perturbative calculations. 
On the other hand, for systems with a scale of about ${\cal O}(1-10)~{\rm GeV}$, 
theoretical uncertainties caused by ``renormalons'' can limit the accuracy of perturbative calculations seriously.

Renormalon \cite{Gross:1974jv,Lautrup:1977hs,`t Hooft:1977jv,Beneke:1998ui} is a notion originating from the infrared (IR) region of certain 
loop diagrams, which is known to cause perturbative coefficients to  diverge 
factorially. 
It makes perturbative expansion an
asymptotic series, indicating that it is impossible to calculate the true value of an observable by perturbative expansion alone.
Renormalons limit achievable accuracy
of a perturbative calculation to be order $(\LQ/Q)^{n}$ with 
an integer $n$, for an observable characterized by a typical 
energy scale $Q(\gg\LQ\sim300~{\rm MeV})$. 
For a system of the electroweak scale, $\LQ/Q\sim 10^{-3}$ would be negligible at present,
while for a system of the bottom or charm quark, $\LQ/Q\sim 0.1$ jeopardizes the precision of the prediction considerably.
In high-precision flavor physics, it is desirable to 
remove the uncertainty due to renormalons from perturbative calculations.

Today,
it has become a standard prescription to use a short-distance quark
mass  (e.g.\ $\msbar$ mass) in various
perturbative QCD calculations,
and this is a way to remove renormalons.
For instance, it was known that the perturbative expansion of
the static QCD potential 
shows a drastically 
divergent behavior due to ${\cal O}(\LQ)$ 
renormalon~\cite{Aglietti:1995tg}. 
When the QCD potential is combined 
with twice of the quark pole mass, the sum constitutes the main
part of the total energy of a heavy quarkonium.
The perturbative series of the pole mass, expressed in terms of 
a short-distance mass,
also shows a divergent behavior due to ${\cal O}(\LQ)$ 
renormalon \cite{Beneke:1994sw,Bigi:1994em}. 
These divergent behaviors cancel out in the combination as the total energy \cite{Pineda:id,Hoang:1998nz,Beneke:1998rk}.
Besides heavy quarkonium observables, the ${\cal O}(\LQ)$  renormalon 
in the $B$ meson (semileptonic)
inclusive decay width is also known to cancel 
out by rewriting the pole mass by a short-distance 
mass \cite{Bigi:1994em,Neubert:1994wq,Ball:1995wa}.
The cancellation of renormalons has improved convergence of the
perturbative series significantly. 
These aspects have been applied successfully in accurate determinations of fundamental physical constants, such as the masses of the heavy quarks \cite{Penin:2014zaa,Kiyo:2015ufa,Beneke:2016oox,Bazavov:2018omf,Peset:2018ria}, some of the Cabibbo-Kobayashi-Maskawa matrix elements \cite{Hoang:1998hm,Hoang:1998ng,Alberti:2014yda}, and the strong coupling constant $\alpha_s$ \cite{Bazavov:2012ka}.

In order to improve accuracy of perturbative calculations further,
it is necessary to eliminate renormalons beyond ${\cal O}(\LQ)$, which requires a theoretical framework that systematically incorporates nonperturbative QCD effects.
The operator product expansion \cite{Wilson:1969zs} (OPE) is suited for this
purpose.
The OPE of an observable (with scale $Q$)
is given by an expansion in $1/Q$, and the expansion coefficients 
are factorized into Wilson coefficients and 
nonperturbative matrix elements. 
Each Wilson coefficient can be calculated as perturbative series,
while each matrix element needs to be evaluated nonperturbatively
and has the size of an integer power of $\LQ$.
Thus, the OPE provides a systematic
double expansion in $\LQ/Q$ and $\alfs(Q^2)$.
Since, however, renormalons in the perturbative series of
the Wilson coefficients induce order $(\LQ/Q)^n$
uncertainties, they obscure systematic improvement of theoretical accuracy
by the double expansion.
It is believed that the uncertainties caused by renormalons cancel out within the framework of OPE, where each matrix element
is believed to cancel
(absorb) the corresponding renormalon in the Wilson coefficients \cite{Mueller:1984vh}.
Hence, by subtracting renormalons from the Wilson coefficients
and absorbing them into the matrix elements, 
each term of the OPE can be made well defined.

There have appeared analyses of the OPEs including cancellation of renormalons beyond 
the ${\cal O}(\LQ)$ renormalon of
the pole mass. 
In the OPE of the QCD potential,
the ${\cal O}(\LQ)$ and 
${\cal O}(\LQ^3)$ renormalons in the
leading-order (LO) Wilson coefficient were subtracted
and this OPE was used to determine $\alfs(M_Z^2)$ precisely \cite{Takaura:2018lpw,Takaura:2018vcy,Ayala:2020odx}.
Also the ${\cal O}(\LQ^4)$ renormalon 
contained in the lattice
plaquette action was subtracted and absorbed into the local gluon condensate \cite{Ayala:2020pxq}.
Many different methods to achieve subtraction of renormalons
from the LO Wilson coefficient have been developed recently 
\cite{Lee:2002sn,Ayala:2019uaw,Ayala:2019hkn,Ayala:2020odx,Ayala:2020pxq,Takaura:2020byt,Hayashi:2020ylq,Hayashi:2021vdq,Hayashi:2021ahf,Benitez-Rathgeb:2022yqb,Benitez-Rathgeb:2022hfj}.
Among them, the FTRS method \cite{Hayashi:2020ylq,Hayashi:2021vdq,Hayashi:2021ahf} 
utilizes the general property that
IR renormalons of the LO Wilson coefficient can be suppressed by taking the
Fourier transform of the Wilson coefficient.

In this paper we propose another renormalon subtraction method
by replacing the
Fourier transform in the FTRS method by the Laplace transform.
In this way we can improve convergence property compared to
the FTRS method without losing its merits.
We 
call this method ``dual space renormalon subtraction''
(DSRS) method, by regarding the Laplace transform and its
inverse transform as transformation between dual spaces
(in analogy to the Fourier transform as transformation
between the position and momentum
spaces).

Subtraction of renormalons is generally scheme
dependent. 
The DSRS and FTRS methods give 
the results in the same scheme as in most other methods, the
principal value (PV) scheme \cite{Lee:2002sn,Ayala:2019uaw,Ayala:2019hkn,Ayala:2020odx,Ayala:2020pxq,Takaura:2020byt}.
These methods provide 
one-parameter integral representations
of the Wilson coefficient,
which are
equivalent to the conventional formula for the PV scheme
expressed by the Borel integral.
A major characteristics, 
as compared to other methods, is the ability to simultaneously separate the contributions of multiple renormalons, i.e., effects of different powers in $\LQ/Q$,
by a simple procedure. 
In principle, the other methods can also separate multiple renormalon contributions, but the separation of renormalons corresponding to 
higher powers of $\LQ/Q$ requires 
multiple steps of calculation to estimate the magnitudes of the renormalons, which complicates the calculation procedure.
Another unique property of the DSRS and FTRS methods is that
the Landau singularity of the running coupling constant, which is
an unphysical singularity, is regularized together with subtraction of
renormalons.
It is a consequence of contour deformation in the dual space
to avoid the
Landau singularity.
This property stabilizes the calculation of the OPE.

In this paper we also elucidate how the renormalon
cancellation takes place in the entire OPE systematically.
To our knowledge this is a first attempt 
in this direction for a general observable.
(Most previous studies of renormalon subtraction focused on
the analysis of the LO Wilson coefficient.)
For this purpose the dual-space integral representation of an
observable provides a useful framework.
We can apply a generalization of the expansion-by-regions (EBR)
method \cite{Beneke:1997zp,Smirnov:2002pj,Jantzen:2011nz} to this integral and investigate the mechanism of renormalon
cancellation.

The EBR method is a technique for evaluating
an integral by expanding the integrand in small parameters.
The method is widely used in evaluation of Feynman loop integrals or in evaluation of
Wilson coefficients (matching coefficients) in various
effective field theories (EFTs).
In this method, first an integral is regularized by analytic 
continuation (dimensional regularization), then
the integral is separated into contributions from different regions
of the integration variables.
The correct expansion of the original integral is obtained
in a sophisticated way,
order by order in powers of a small parameter
(ratio of external mass scales).

To investigate the properties of the DSRS method 
in detail, we apply the
method to a solvable model, the $O(N)$ nonlinear $\sigma$ model
in two dimensions in the large $N$ limit.
It can be regarded as a
toy model which mimics QCD. 
In this model, various observables can be computed 
exactly, and their OPEs can also be computed.
We show in some examples that the dual transform removes all the IR renormalons
as well as IR divergences of the Wilson coefficients.
This leads to a systematic understanding of renormalon
cancellation.
We demonstrate by a simulation study imitating the QCD case
that the DSRS method can be used to determine 
nonperturbative parameters accurately.

Finally we discuss how to apply the DSRS method to the
QCD case.
We discuss possible complications as well as solutions to them.
Nevertheless, by construction of the DSRS method, we expect the
application to be mostly
straightforward.

The paper is organized as follows.
In Sec.~2 we explain our basic idea how we understand
the OPE and cancellation of renormalons.
In Sec.~3 we explain the framework of the dual space approach.
In Sec.~4 we apply the DSRS method to the $O(N)$ nonlinear $\sigma$ model and
study the OPE of some observables.
In Sec.~5 we discuss application of the DSRS method to QCD
and a remaining problem.
Sec.~6 is devoted to summary and conclusions.
Details are explained in Appendices.
In App.~A we show the relation between the DSRS and FTRS methods.
In App.~B we explain how to resum UV renormalons.
The proof of the EBR method in the dual space approach
is given in App.~C.
Details of the calculations for the $O(N)$ nonlinear $\sigma$ model 
are given in App.~D.

\section{Basic idea}
\label{secBasicIdea}

To begin with, we explain our basic idea on
how renormalons appear and cancel
in the OPE, taking the
static QCD potential as an example.

Consider the static QCD potential defined nonperturbatively,
e.g., as computed by lattice QCD.
It is well defined nonperturbatively both in the
position space and momentum space.
$V(r)$ is related to $\widetilde{V}(q)$ by Fourier transform as
\begin{align}
V(r)
&=\int \frac{d^3 \vec{q}}{(2\pi)^3}\, e^{i\vec{q}\cdot\vec{r}}
\,\, \widetilde{V}(q)
\,.
\label{FT-V}
\end{align}
We focus on the short-distance region $1/r \gg \LQ$,
while we neglect effects from the ultra-soft scale
for simplicity.
We divide Eq.~\eqref{FT-V} in an
EBR manner by
employing dimensional regularization with
$d=3-2\epsilon$.
\begin{align}
V(r)
&=
\lim_{d \to 3}
\Biggl\{
\underbrace{
\int  \!\!\! \frac{d^d \vec{q}}{(2\pi)^d}\, e^{i\vec{q}\cdot\vec{r}}
\,\Bigl[ \widetilde{V}(q) \Bigr]_{\rm OPE}
}_{\begin{array}{c}
{\displaystyle V_{\rm hard}(r)}\\
\text{\rotatebox{90}{$\in$}}
\end{array}}
+
\underbrace{
\int \!\!\! \frac{d^d \vec{q}}{(2\pi)^d}\, 
\Bigl[1+i\vec{q}\!\cdot\!\vec{r}
+\frac{1}{2}(i\vec{q}\!\cdot\!\vec{r})^2 + \dots
\Bigr]
\widetilde{V}(q)
}_{\begin{array}{c}
{\displaystyle V_{\rm soft}(r)}\\
\text{\rotatebox{90}{$\in$}}
\end{array}}
\Biggr\}
\,.
\label{generalEBR-V(r)}
\\
&
~~~~~~~~~~~~~~~~
\pm i N_n (\LMS r)^n/r
~~~~~~~~
\stackrel{\rm cancel}{\Longleftrightarrow}
~~~~~~~~
\mp i N_n (\LMS r)^n/r
\nonumber
\end{align}
On the right-hand side (RHS), we divide $V(r)$ into the contribution
from the hard region, $q=|\vec{q}|\simgt 1/r$,
and that from the soft region, $q\simlt\LQ\ll 1/r$.
These contributions are denoted as $V_{\rm hard}(r)$ and
$V_{\rm soft}(r)$, respectively.
In the hard region, since $q\gg\LQ$, we replace
$\widetilde{V}(q)$ by its high-energy expansion,
namely by the OPE, $[\widetilde{V}(q)]_{\rm OPE}$.
In the soft region, instead we expand the factor
$e^{i\vec{q}\cdot\vec{r}}$ in $q$ $(\ll 1/r)$,
while $\widetilde{V}(q)$ should be evaluated nonperturbatively.
We obtain a series expansion in $r$,
$V_{\rm soft}(r)=\sum_n d_n r^n$,
with the expansion
coefficients $d_n$ defined nonperturbatively.
As is the case of the usual EBR of a Feynman integral,
we expect that $V_{\rm hard}(r)$ and
$V_{\rm soft}(r)$ are given by
expansions in $r$ (with $\log r$ corrections in the former).
We also expect that
the expansion coefficients in each of $V_{\rm hard}(r)$
and $V_{\rm soft}(r)$
contain IR and UV divergences as $1/\epsilon$ poles,
while in the sum $V_{\rm hard}(r)+V_{\rm soft}(r)$
the divergences cancel since $V(r)$ is finite
as $\epsilon \to 0$.
In this way, the EBR gives the OPE of $V(r)$ in $r$.

Similarly to the appearance of $1/\epsilon$ poles,
there also appear imaginary part in the form
$\pm i N_n (\LMS r)^n/r$
in each of $V_{\rm hard}(r)$
and $V_{\rm soft}(r)$ for $n=1,3,5,\dots$.
The imaginary part in the Wilson coefficients of the OPE
originate from renormalons.
(This feature is based on the large-$\beta_0$ approximation, and for simplicity we neglect corrections to this approximation in this section.\footnote{
Namely, we 
neglect corrections to the integer power behavior in $\LQ$ of the
renormalons.
})
The imaginary part should
cancel in the sum $V_{\rm hard}(r)+V_{\rm soft}(r)$
since $V(r)$ is real.\footnote{
The $r$-independent imaginary part (by the ${\cal O}(\LQ)$ renormalon) in the leading
Wilson coefficient of $V_{\rm hard}(r)$
also cancels
against the imaginary part 
in $V_{\rm soft}(r)$
in the nonperturbative definition of $V(r)$.
This may look different from the cancellation of the 
${\cal O}(\LQ)$ renormalon
in the perturbative expansion of $V(r)$
against the corresponding renormalon in twice of the pole mass
(see Sec.~\ref{sec1}).
Note, however, that the nonperturbative definition of $V(r)$
via Wilson loop includes
the self-energies of
the static charges, which are evaluated to be
zero to all orders of perturbative
expansion but nonzero nonperturbatively.
}

In short, the imaginary part of $V_{\rm hard}(r)$ appears as follows.
IR renormalons in the leading order (LO) Wilson coefficient of
$[\widetilde{V}(q)]_{\rm OPE}$ are suppressed,
as shown in ref.~\cite{Sumino:2020mxk}.
Therefore, the Wilson coefficient in $q$ space
can be calculated accurately in series 
expansion in $\alfs(q^2)$.
Then 
the IR renormalons in the regularized LO Wilson coefficient
of $V_{\rm hard}(r)$
can be separated
as imaginary part, using the 
contour integral surrounding the Landau
singularity of $\alfs(q^2)$ in the complex $q$-plane
(FTRS method)~\cite{Hayashi:2020ylq,Hayashi:2021vdq,Hayashi:2021ahf}.
The renormalon subtraction
in this way agrees with the conventional PV prescription
in the Borel resummation method in the
all-order limit of the perturbative expansion.

The higher-order operators and associated
Wilson coefficients in the OPE of $V(r)$ or $\widetilde{V}(q)$
can be obtained using the potential nonrelativistic QCD EFT 
\cite{Brambilla:1999qa,Brambilla:2004jw}
by matching to full QCD.
Hence, 
we expect that we can apply
the same technique also to 
the subleading Wilson coefficients
in $[\widetilde{V}(q)]_{\rm OPE}$
(although we need to solve problems
caused by the ultrasoft scale).

Cancellation of the imaginary part is a general feature
which is guaranteed by validity of the generalized
EBR method.
However, explicit confirmation of the cancellation
is difficult in the case of QCD, since $V_{\rm soft}(r)$
is difficult to calculate in dimensional regularization.
The cancellation can be checked explicitly
in a solvable model such as the 2D $O(N)$ nonlinear $\sigma$ model.

The EBR method used here and throughout this paper is
different from the ordinary EBR method used for expanding loop integrals.
We note that we use the EBR to the most external momentum integral only,
and we do not apply EBR to the construction of $\widetilde{V}(q)$. 
We use the terminology ``hard'' and ``soft'' with respect to this
most external momentum in Eq.~\eqref{generalEBR-V(r)}.
We will give a detailed explanation in the following sections.

In the following sections we will apply the
above idea to general observables
using the DSRS method.
We will replace the Fourier transform by
the dual transform, by which we separate renormalons more
efficiently.
We will explain how the imaginary part arise in the conventional
approach and also in the dual space approach.


\section{Framework of dual space approach}
\label{sec2}

In this section
we explain the framework of the dual
space approach based on the Laplace transform.
In Sec.~\ref{ss:OPE},
we review the conventional treatment of
IR renormalons and the OPE.
In Sec.~\ref{ss:DS}
we introduce the  dual space
in which IR renormalons are suppressed.
We use it to derive an expression of
the renormalon-subtracted LO Wilson coefficient.
In Sec.~\ref{ss:MB1}
we construct the OPE of a general observable 
in the dual space approach and
explain how renormalons emerge and are canceled.

\subsection{OPE, IR renormalons and regularized Wilson coefficient}
\label{ss:OPE}

We consider a dimensionless observable $S(Q^2)$ 
which is renormalization group (RG) invariant
and characterized by a single hard scale $Q$ ($\gg\LQ$).
Its OPE is expressed as\footnote{
For many observables of our interest, the LO nonperturbative
matrix element is one, $\langle{{\cal O}_0}\rangle=1$, or equivalently,
the LO Wilson coefficient is identical to
perturbative evaluation of the observable.
We consider 
those cases in this paper.
}
\be
\Bigl[ S(Q^2)\Bigr]_{\rm OPE}
=C_0(Q^2)+C_1(Q^2)\frac{\langle{{\cal O}_1}\rangle}{Q^{d_1}}+\cdots,
\label{SOPE}
\ee
where $C_0, C_1$ are the Wilson coefficients
and $\langle{{\cal O}_1}\rangle$ is the matrix element of 
an operator $\mathcal{O}_1$ of mass dimension
$d_1$.
The Wilson coefficients can be computed by perturbative calculation,
and in particular, we express the LO Wilson coefficient as
\begin{align}
\big[C_0(Q^2)\big]_{\rm PT}
=\sum_{n=0}^\infty \alpha_s(\mu^2)^{n+1} c_n(L_Q)
=\sum_{n=0}^\infty \alpha_s(Q^2)^{n+1} c_n(0)
\label{eq:c0pt1}
\,,
\end{align}
where $\alfs(\mu^2)$ is the coupling constant renormalized
at the renormalization scale $\mu$;
the coefficients $c_n(L_Q)$ are polynomials in $L_Q=\log\big(\mu^2/Q^2\big)$.
In the second equality we set $\mu=Q$ using the RG invariance.

In asymptotic-free theories such as QCD,
naive perturbative series
has IR renormalons,
namely, Eq.~\eqref{eq:c0pt1} is a divergent series 
where the coefficient grows factorially, $c_n(0) \sim n!$.
The conventional method to 
regularize such a divergent series is to use
the Borel resummation method.
The Borel transform of Eq.~\eqref{eq:c0pt1} is defined as
\be
B_{C_0}(u)=\sum_{n=0}^\infty\frac{c_n(L_Q)}{n!}(u/b_0)^n
\label{BorelTr-C0-3.3}
\,.
\ee
Here, $b_0$ denotes the  one-loop coefficient of the beta function.
(We adopt the convention where $b_0=(11-2n_f/3)/(4\pi)$
in the case of QCD.)
Due to the acceleration of convergence by $1/n!$,
this series in $u$ has a finite radius of convergence,
and the IR renormalons are
seen as singularities on the positive real axis 
in the complex $u$-plane.\footnote{
We neglect effects of singularities other than IR renormalons.
For example, instanton effects are also known to generate
singularities on the positive real axis
(located far from the origin compared to the LO IR renormalon),
but we neglect their effects.
}
In the leading-logarithmic (LL) or large-$\beta_0$ approximation, each renormalon singularity is 
a pole, while including subleading logarithmic corrections,
it becomes a branch point.
We denote the positions of the IR
renormalon singularities as $u=u_*$($>0$),
which make up an infinite sequence of integers or half-integers
that depends on the observable.
In particular, the most dominant renormalon, which produces the fastest divergent growth,
corresponds to 
the singularity closest to $u=0$.
The regularized LO Wilson coefficient is constructed as
\be
\big[C_0(Q^2)\big]_\pm^{\rm Borel}=\frac{1}{b_0}\int_{C_\pm}du\,B_{C_0}(u)\,e^{-u/(b_0\alpha_s(\mu^2))},
\label{Cpm-Borel}
\ee
where
$\int_{C_\pm}du = \int_{0_\pm i0}^{\infty_\pm i0}du$.\footnote{
For a convergent series without IR renormalons,
Eq.~\eqref{Cpm-Borel} is simply the inverse Borel transform.
}
The IR renormalon singularities are avoided by deforming the integration contour 
infinitesimally above/below the positive real axis.

\begin{figure}[tbp]
\centering
\includegraphics[width=6cm]{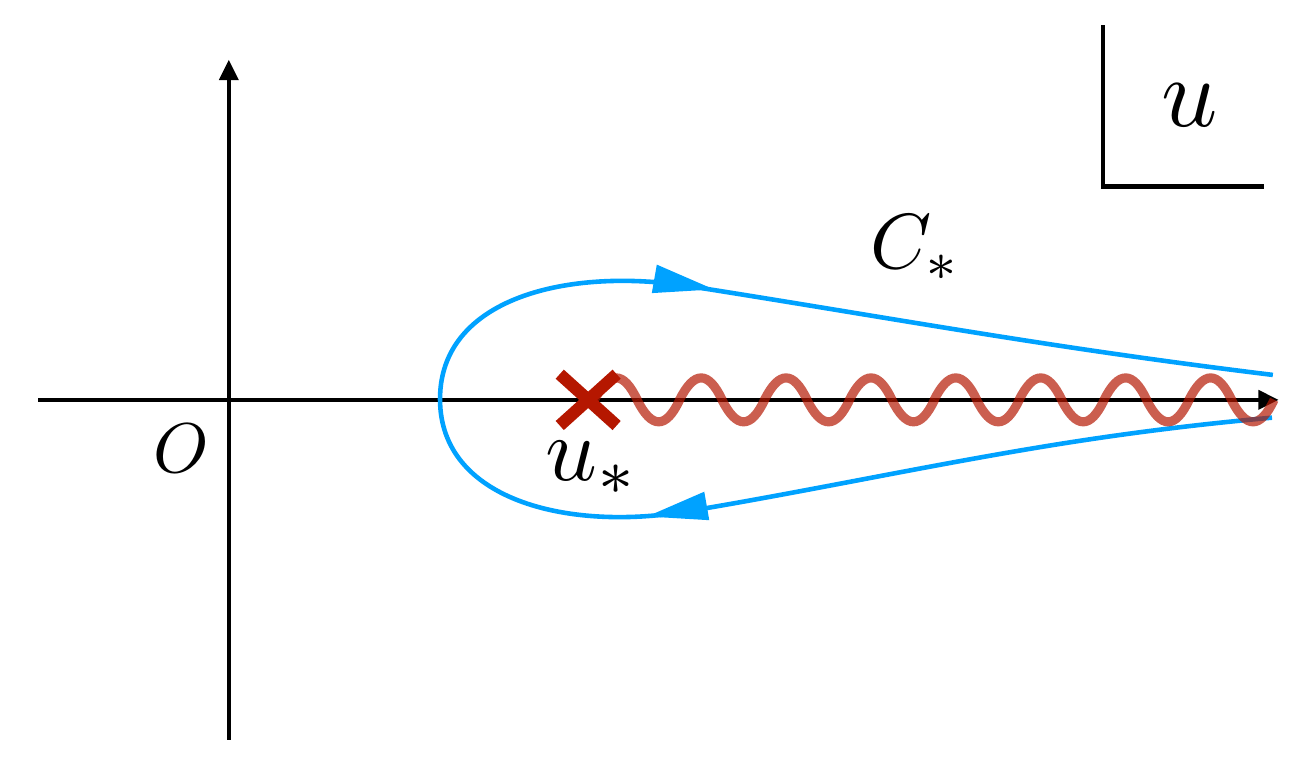}
\caption{\small
Contour $C_*$ for the imaginary part of the regularized Wilson coefficient
in Eq.~\eqref{ren-form-fromOPE}.
}
\label{Fig-Cstar}
\end{figure}

Since $\big[C_0(Q^2)\big]_\pm^{\rm Borel}$
are complex conjugate to each other,
their difference
$2i \delta C_0\equiv\big[C_0(Q^2)\big]_+^{\rm Borel}-\big[C_0(Q^2)\big]_-^{\rm Borel}$
is pure imaginary.
The contribution of each renormalon to $\delta C_0$
can be calculated
in terms of the residue of the renormalon or 
the integral around the branch cut:
\begin{align}
\delta C_0^{u_*}&=\frac{1}{2 i b_0}\int_{C_*}du\,B_{C_0}(u)\,e^{-u/(b_0\alpha_s(\mu^2))}
\nonumber\\
&=
\frac{\pi \, N_{u_*}\, {u_*^{1+\nu_{u_*}}}}{b_0\,\Gamma(1+\nu_{u_*})}
\left(\frac{\Lambda_{\rm \overline{MS}}^2}{Q^2}\right)^{\! u_*}\!\!\!
(b_0\alpha_s(Q^2))^{\gamma_0/b_0}
\bigg(1+\sum_{n=0}^\infty {s}_n\alpha_s(Q^2)^{n+1}\bigg)
\,.
\label{ren-form-fromOPE} 
\end{align}
$C_*$ is shown in Fig.~\ref{Fig-Cstar}.
In the second line, the form of $\delta C_0^{u_*}$ is
determined 
assuming that the imaginary part is
canceled by the corresponding term of OPE,
Eq.~\eqref{SOPE}
(in the case that there is a unique operator 
with dimension $2u_*$)~\cite{Mueller:1984vh}.
Apart from the overall normalization $N_{u_*}$, the parameters $b_0$, $\nu_{u_*}$, 
$\gamma_0$ and the expansion
coefficients $s_n$ can be calculated in perturbation theory; 
$\Lambda_{\rm \overline{MS}}$ denotes the integration constant
of the running coupling constant in the $\msbar$ scheme.

Thus, renormalons generate imaginary part to
the regularized Wilson coefficient,
and its sign depends on how
the renormalon singularities are avoided along the integral contour
in the $u$ plane:
\be
\big[C_0\big]_\pm^{\rm Borel}
= \big[C_0\big]_{\rm PV} \pm i \delta C_0
\,, ~~~~~
[C_0]_{\rm PV}= \frac{1}{b_0}\int_{0,\rm PV}^{\infty}
du \, e^{-u/(b_0\alfs)}  B_{C_0}(u)\,,
\ee
where PV stands for the principal value integration
or the average of the $C_\pm$ integrals.

There is scheme dependence in how to subtract renormalons 
from the Wilson coefficient.
A conventional prescription,
the ``principal value (PV) prescription," is 
to define a renormalon-subtracted Wilson coefficient by $[C_0]_{\rm PV}$.
By definition, the renormalon contributions are minimally subtracted from 
$C_0(Q^2)$.

\subsection{Dual transform of Wilson coefficient and
renormalon suppression
}
\label{ss:DS}

Next we present an alternative expression of
the regularized Wilson coefficient using a dual transform.
Our argument is based on the formula of the inverse Laplace transform, 
\be
\frac{1}{2\pi i}\int_{t_0-i\infty}^{t_0+i\infty} \!\!\!\!dt\,e^{tp^2}t^{u}
=\frac{1}{(p^2)^{1+u}}\frac{1}{\Gamma(-u)}
~~~;~~~t_0>0
\,.
\label{inv-Lap-tr}
\ee
We introduce two parameters $(a,u')$
and define the LO Wilson coefficient in the dual space as
\be
\tilde{C}_0(p^2)
=\frac{1}{2\pi i}\int_{t_0-i\infty}^{t_0+i\infty} \!\!\!\!dt\,e^{tp^2}t^{au'}
C_0(t^{-a}) \,,
~~~~~~~
t=Q^{-2/a} \,.
\label{eq:dualLOW}
\ee
It transforms the $\alfs$ expansion in $Q^2$ space Eq.~\eqref{eq:c0pt1}
to the $\alfs$ expansion in a dual space ($p^2$ space).
This is regarded as a dual transformation,
similarly to 
the Fourier transform as a transformation 
from the coordinate space to the momentum space.

With a proper choice of $(a,u')$, the factorial growth
of the series due to IR renormalons can be suppressed in the dual space.
This can be seen as follows.
We can apply the (regularized) Borel resummation to
both sides of Eq.~\eqref{eq:dualLOW} with respect to
the series expansions in $\alfs(\mu^2)$.
$C_0$ on the RHS
is replaced by $\big[C_0\big]_\pm^{\rm Borel}$,
and the left-hand side (LHS) is replaced by 
$\big[\tilde{C}_0\big]_\pm^{\rm Borel}$.
Then we take the difference of the $\pm$ contributions
and express the RHS as the sum of the contributions
from all the singularities $u_*$.
Thus, by replacing 
$C_0(Q^2)$ by 
$\delta C_0^{u_*}(Q^2)\approx\big(\LMS^{2}/Q^{2}\big)^{u_*}$
in Eq.~\eqref{eq:dualLOW},
we obtain 
\bea
\delta \tilde{C}_0^{u_*}(p^2)
&\approx&\frac{1}{2\pi i}\int_{t_0-i\infty}^{t_0+i\infty} \!\!\!\!\!\!dt\,e^{tp^2}t^{au'}\big(\LMS^2\,t^{a}\big)^{u_*}
\label{eq:suppIRrenom0}
\\
&=&\frac{1}{(p^2)^{1+au'}}\frac{1}{\Gamma(-a(u_*+u'))}
\left(\frac{\LMS^2}{p^{2a}}\right)^{u_*}
\label{eq:suppIRrenom1}
\,.
\eea
Eq.~\eqref{eq:suppIRrenom1} is zero if 
$u_*+u'=n/a$ for $n\in\mathbb{Z}_{\ge 0}$.
Namely, we can choose $(a,u')$ such that there no longer exist singularities
in the Borel plane which induce factorial growth
of $\tilde{C}_0(p^2)$ if the imaginary part 
of  $\big[C_0(Q^2)\big]_\pm^{\rm Borel}$
are of the form
$(\LMS^2/Q^2)^{n/a-u'}$.
Thus, IR renormalons of $\tilde{C}_0(p^2)$ can be suppressed.
It is possible to suppress more than one IR renormalons simultaneously 
with an appropriate choice of $(a,u')$~\cite{Hayashi:2020ylq,Hayashi:2021vdq,Hayashi:2021ahf}.
In principle, 
it is also possible to
take into account 
the higher-order $\alpha_s(Q^2)$ corrections to 
the simple power of $\LMS^2/Q^2$
and anomalous dimension effects
order by order. (See the end of this subsection.)
Hence, the perturbative expansion of $\tilde{C}_0$ can be regarded as a convergent series.

While $\big[ \tilde{C}_0 \big]_\mathrm{PT}$ is a convergent series,
$\big[ C_0 \big]_\mathrm{PT}$ is not, due to the IR renormalons.
The difference emerges as follows.
\be
\big[\tilde{C}_0(p^2)\big]_{\rm PT}
=\frac{1}{2\pi i}\int_{t_0-i\infty}^{t_0+i\infty} \!\!\!\!dt\,e^{tp^2}t^{au'}\big({\mu^2}
t^a\big)^{\hat{H}}\sum_{n=0}^\infty c_n(0)\alpha_s(\mu^2)^{n+1}
\,,
\label{til-C0-PT}
\ee
where, using the RG invariance,
we have rewritten
Eq.~\eqref{eq:c0pt1}  as
\begin{align}
\big[C_0(Q^2)\big]_{\rm PT}
=\bigg(\frac{\mu^2}{Q^2}\bigg)^{\hat{H}}
\sum_{n=0}^\infty \alpha_s(\mu^2)^{n+1} c_n (0)\,,
\label{C0PT}
\end{align}
with
\be
\hat{H}
=-\beta(\alfs)\,\frac{\partial}{\partial\alpha_s(\mu^2)}
\,,
~~~~~~~~
\beta(\alfs)
=\mu^2\frac{d}{d \mu^2}\alfs(\mu^2)
=-\sum_{i=0}^\infty b_i\alfs^{i+2} (\mu^2) 
\,.
\label{opHinRGE}
\ee
Integrating over $t$ we have
\begin{align}
\big[\tilde{C}_0(p^2)\big]_{\rm PT}
&=\frac{1}{(p^2)^{1+au'}}\,
e^{\hat{H}\,\log(\mu^2/p^{2a})}\,\frac{1}{\Gamma(-a(\hat{H}+u'))}\,
\sum_{n=0}^\infty c_n(0)\alpha_s(\mu^2)^{n+1}
\non
&=\frac{1}{(p^2)^{1+au'}}\,e^{\hat{H}\,\log(\mu^2/p^{2a})}\,
\sum_{n=0}^\infty \tilde{c}_n(0)\alpha_s(\mu^2)^{n+1}
\,,
\label{til-C0-PT2}
\end{align}
where
$\tilde{c}_n(0)$'s are computed by comparing $\alfs(\mu^2)$ expansion 
(using the expansion in $\hat{H}$) of 
the first and second lines.
We can see that the coefficients of the $\alfs(\mu^2)$ expansion 
of Eq.~\eqref{til-C0-PT2} have $\log^n(\mu^2/p^{2a})$ dependence.
The Laplace integral of such logarithms in the region 
$p^{2a}\simlt \LMS^2$ gives a factorial behavior $\sim n!$, 
which is the origin of the IR renormalons in $C_0$.
Noting that $\tilde{C}_0$ is RG invariant, we can resum the logarithms 
by setting $\mu^2=p^{2a}$ in Eq.~\eqref{til-C0-PT2}:
\be
\big[\tilde{C}_0(p^2)\big]_{\rm PT}
=\frac{1}{(p^2)^{1+au'}}\,
\sum_{n=0}^\infty \tilde{c}_n(0)\alpha_s(p^{2a})^{n+1}
\label{tildeC0_PT}
\,.
\ee
As a result, 
the higher power logarithms are 
turned into the behavior of the 
running coupling constant $\alpha_s(p^{2a})$ around the Landau singularity,
which becomes an alternative source of renormalons upon considering the Laplace
integral.
The expansion of $\tilde{C}_0(p^2)$ in $\alpha_s(p^{2a})$ is also a convergent series for a sufficiently small $\alpha_s(p^{2a})$. 
In practical calculations, the original series Eq.~\eqref{C0PT} is known only
up to a certain order $n\le k$.
Then we compute $\tilde{c}_n(0)$ also up to $n=k$.
Thus, using the running coupling constant 
of the N$^k$LL approximation and truncating $\tilde{C}_0(p^2)$ at
order 
$\alfs(p^{2a})^{k+1}$, one can compute a good approximation 
(at the level of N$^k$LL)
of the true $\tilde{C}_0(p^2)$, which is defined by summing up the series to all orders.

The original Wilson coefficient $C_0(Q^2)$
can be reconstructed by the Laplace transform of $\tilde{C}_0(p^2)$.
As a result of the resummation described above, the singularity of the running coupling constant arises in the integral path along 
the real axis of the Laplace integral.
Choosing the same integration contour $C_\pm$ as Eq.~\eqref{Cpm-Borel},
the regularized Wilson coefficient $\big[C_0(Q^2) \big]_\pm$ is given by
\begin{align}
\big[C_0(Q^2)\big]_\pm
&=t^{-au'}\int_{C_\mp}dp^2\,e^{-tp^2}\tilde{C}_0(p^2)
\label{eq:LT}
\\
&=\big[C_0(Q^2)\big]_{\rm PV}\pm \,i\,\delta C_0(Q^2),
\end{align}
\be
\big[C_0(Q^2)\big]_{\rm PV}=t^{-au'}\int_{0,\rm PV}^\infty\!\!\!\! dp^2\,e^{-tp^2}\tilde{C}_0(p^2),\quad
\delta C_0(Q^2)=\frac{t^{-au'}}{2i}\int_{C_--C_+}\!\!\!\!\!\!\!\!\!\!dp^2\,e^{-tp^2}\tilde{C}_0(p^2)
\label{C0_PV+deltaC0}
\,,
\ee
where PV stands for the real part of the integration.
It can be proven that $\big[C_0(Q^2)\big]_{\pm}$ given here 
agrees with that defined via the Borel resummation
in the previous subsection. (See below.)
Keeping the integral value,
the integral path $C_\mp$ can be deformed 
away from the Landau singularity in the right-half plane, such that
$\bigl{|}\alpha_s(p^{2a})\bigr{|}$ is sufficiently small along the entire
integral path.\footnote{
We neglect effects of the singularities which possibly exist in 
the right-half plane (${\rm Re}\,p^2>0$) other than the Landau singularity.
This would be reasonable for observables defined in Euclidean momentum
region (e.g., Adler function, QCD potential).
At the level of N$^k$LL approximation in QCD ($k\le 4$), singularities
which affect our present argument
do not exist; see App.~C of ref.~\cite{Hayashi:2021vdq} for details.
}
Hence, we obtain a good estimate of $\big[C_0(Q^2)\big]_{\rm PV}$
and $\delta C_0(Q^2)$ using
the series expansion of $\tilde{C}_0(p^2)$.
In particular, we call this method of computing $\big[C_0(Q^2)\big]_{\rm PV}$
dual space renormalon subtraction (DSRS) method.


As compared to 
the FTRS
method~\cite{Hayashi:2020ylq,Hayashi:2021vdq,Hayashi:2021ahf}, the DSRS method is more efficient.
The Fourier transform in the FTRS method
generates artificial UV renormalons, which
worsen the convergence of the series in
the Fourier space.
In contrast the dual transform suppresses IR renormalons
without generating
artificial UV renormalons.
It turns out that the DSRS method 
is equivalent to the FTRS method
after resummation of the artificial UV renormalons in the latter.
We give the proof in App.~\ref{app:RelToFTRS}.
Once this is understood, we can prove
the equivalence of $\big[C_0(Q^2)\big]_{\pm}$ calculated by
the Borel resummation and by the DSRS method
through the equivalence of the former and that by
the FTRS method which is shown in App.~C of ref.~\cite{Hayashi:2021vdq}.
Similarly, we can incorporate
the corrections to the
integer power forms proportional to
$ (\LQ/Q)^n$ of the imaginary part,
given by the anomalous dimension and expansion in $\alfs(Q^2)$,
into the DSRS method.
This can be achieved
by using the corresponding formula for the
FTRS method, given in App.~B of ref.~\cite{Hayashi:2021vdq}.
(See, however, the discussion at the end of Sec.~\ref{sec5}.)

\subsection{OPE in dual space approach}
\label{ss:MB1}

We explain how the OPE of an observable
is derived in the dual space approach.
Consider an observable which is
calculated nonperturbatively
(using lattice calculation, for example).
Since it does not include perturbative series, 
it includes no renormalon ambiguities.
Hence, its dual transform and inverse transform can
be calculated without regularization of the integrals.
\begin{align}
&
S(Q^2)
=
t^{-au'}\int_{0}^\infty dp^2\,e^{-tp^2}\tilde{S}(p^2) \,,
~~~~~~~
t=Q^{-2/a}
\,,
\label{exact-LaplaceInt}
\\ 
\Longleftrightarrow
~~~
&
\tilde{S}(p^2)
=\frac{1}{2\pi i}\int_{t_0-i\infty}^{t_0+i\infty} \!\!\!\!dt\,e^{tp^2}t^{au'}
S(t^{-a}) \,,
~~~~~~~
t_0>0
\,.
\label{eq:exactDT}
\end{align}
Below we assume that the above integrals are well defined.

In the case $Q\gg\LQ$, we obtain the OPE of $S(Q^2)$
by evaluating the $p^2$ integral
in Eq.~\eqref{exact-LaplaceInt}
by expansion in $1/Q$
using a generalized EBR method.
We divide the integral into the contribution from the
hard region, $p^2\simgt t^{-1}=Q^{2/a}$, and that from
the soft region, $p^2\ll t^{-1}$, after regularizing
the integral.
\begin{align}
\Bigl[ S(Q^2)\Bigr]_{\rm OPE}
=\lim_{\delta \to 0} \biggl\{
&
t^{-au'}\int_{C_\mp} dp^2\,p^{-2\delta}\,e^{-tp^2}
\left[\tilde{S}(p^2)\right]_{\rm OPE}
\nonumber\\&
+
t^{-au'}\int_{C_\mp} dp^2\,p^{-2\delta}\,
\left[
1-tp^2+\frac{1}{2}(tp^2)^2-\dots
\right]
\tilde{S}(p^2)
\biggr\}
\,.
\label{DSOPE-general}
\end{align}
In the hard region, $\tilde{S}(p^2)$
is replaced by its OPE, namely expansion in $1/p^2$
and $\alfs(p^{2a})$.
In the soft region, $e^{-tp^2}$ is expanded in $p^2$.
The IR and UV divergences which originate from the
expansions
are regularized by introducing
$\delta$ analogously to the dimensional regularization,
while the Landau singularity included in 
$\bigl[\tilde{S}(p^2)\bigr]_{\rm OPE}$
is regularized by deforming 
the integral contour to $C_\mp$.
(Or, we can deform the contour
further away from the Landau singularity without changing the integral values.)
The same contour $C_\mp$
is chosen in both hard and soft contributions.

Each term of
$\bigl[\tilde{S}(p^2)\bigr]_{\rm OPE}
=\sum_i \tilde{C}_i(p^2) \langle O_i\rangle /p^{n_i}
$
is uniquely determined by the dual transform of each term of
$\bigl[ S(Q^2)\bigr]_{\rm OPE}$
given by Eq.~\eqref{SOPE}.\footnote{
We assume that the OPE is equivalent to the
high-energy expansion of $S(Q^2)$ or $\tilde{S}(p^2)$.
}
Hence, the Wilson coefficients $\tilde{C}_i(p^2)$
can be calculated in
perturbative expansion
following the same calculational procedure
as in the previous subsection, and there are no
IR renormalons in the perturbative series.
In this derivation of
$\bigl[ S(Q^2)\bigr]_{\rm OPE}$, 
we evaluate $\tilde{C}_i(p^2)$ in expansion in $\alfs(p^{2a})$,
which generates
the Landau singularity in the integrand.
Convergence of the first term of 
Eq.~\eqref{DSOPE-general} follows from the fact that
the integral contour can be deformed,
such that it is sufficiently
away from the Landau singularity
and the integrand converges as we include higher
order terms of
perturbative expansion in $\alfs(p^{2a})$.

Since the original integral 
in Eq.~\eqref{exact-LaplaceInt}
has no singularities and is well defined,
the IR and UV divergences (regularized by $\delta$)
as well as imaginary part must
cancel in the sum of the hard and soft contributions.
This property is guaranteed by the validity
of the generalized EBR method.
We will prove the validity in the cases of the
$O(N)$ nonlinear $\sigma$ model and QCD (in N$^k$LL approximation) in
the following sections.
We will see that the sign of the imaginary part of $\delta$
must be correlated with the choice of the regularized
contour $C_\mp$ in order to guarantee
the validity of the EBR method.

The evaluation of the soft contribution,
the second term of Eq.~\eqref{DSOPE-general},
involves nonperturbative evaluation of $\tilde{S}(p^2)$,
which is possible only in limited cases.
We will confirm
desired properties for some examples in
the $O(N)$ nonlinear $\sigma$ model.
On the other hand, we can always reduce the nonperturbative
information to a few nonperturbative
parameters (matrix elements of local operators)
if we use only the first few terms of the OPE.
This aspect is useful in practical applications of the OPE
for precise theoretical predictions of observables.


\section{\boldmath
OPEs in nonlinear $\sigma$ model}

In this section we demonstrste usefulness of the dual space approach 
through a study
of the OPEs in the 2D $O(N)$ nonlinear $\sigma$ model.
In Sec.~\ref{sec3:model} we briefly explain the model and fix notation.
In Secs.~\ref{sec3:gradient-flow} and \ref{sec3:simulation-study} we analyze the OPE of a simple observable
in detail.
We confirm emergence and cancellation of renormalons in the dual space
approach (Sec.~\ref{sec3:gradient-flow}).
It is followed by a simulation study, where we see that
accuracy of theoretical prediction can be improved
by the DSRS method (Sec.~\ref{sec3:simulation-study}).
In Sec.~\ref{sec3:SEofSigma} we investigate the OPE of the self-energy of $\sigma$
as another example.

\label{sec3}
\subsection{The model}
\label{sec3:model}

We consider the $O(N)$ nonlinear $\sigma$ model in two dimensions
in the large-$N$ limit~\cite{Bardeen:1976zh}.
The partition function of the model is given by
\be
Z=\int{\cal D}\sigma{\cal D}\alpha\exp\bigg[-\frac{1}{2g_0}\int d^2x\bigg\{\partial_\mu\sigma^a\partial_\mu\sigma^a+{\alpha}\bigg(\sigma^a\sigma^a-N\bigg)\bigg\}\bigg],
\ee
where $g_0$ is the bare coupling constant.
The Lagrange multiplier $\alpha(x)$ is introduced 
to take into account the constraint $\sum_{a=1}^N\sigma^a\sigma^a=N$.
This model is solvable: it is an asymptotically free theory
and a mass gap is generated nonperturbatively.
(The $O(N)$ symmetry is unbroken.)
The mass gap is
given by the vacuum expectation value of $\alpha$ as
\be
\langle\alpha\rangle=m^2\equiv \mu^2\,\exp\big[-4\pi/g(\mu^2)\big]
\label{eq:alphaVEV}
\,,
\ee
where $g(\mu^2)$ denotes the coupling constant in the $\overline{\rm MS}$ scheme
renormalized at the renormalization scale $\mu$.
For convenience, we use the rescaled coupling constant 
below:
\be
\bar{g}(\mu^2)=\frac{g(\mu^2)}{4\pi}=\frac{1}{\log(\mu^2/m^2)}
\,.
\ee
It diverges at the Landau pole $\mu^2=m^2$.
The propagater of $\delta\alpha\equiv\alpha-\langle\alpha\rangle$ is given by
\be
D_\alpha(p^2)
=
\int d^2x \,e^{-ip\cdot (x-y)}\,
\langle \delta \alpha(x) \delta\alpha(y) \rangle
=\frac{1}{N}\,
\frac{4\pi\sqrt{p^2(p^2+4m^2)}}{\log\bigg[\frac{\sqrt{p^2+4m^2}+\sqrt{p^2}}{\sqrt{p^2+4m^2}-\sqrt{p^2}}\bigg]}
\,.
\label{Dalpha}
\ee
We investigate the OPEs of observables in this model using the
dual space approach.

\subsection{\boldmath Local condensate of $\delta\alpha$ regularized by 
gradient flow}
\label{sec3:gradient-flow}
The first example is a vacuum expectation value of
a composite operator of $\delta \alpha(x)$ regularized
by gradient flow,
\begin{align}
I(t)
&\equiv Nt^2\cdot
\langle \delta \alpha(x) \,e^{t\,\Box_x}\delta\alpha(x) \rangle
=Nt^2\int_0^\infty{dp^2}\, 
\frac{D_\alpha(p^2)}{4\pi}\,e^{-tp^2}
\,,
\label{It-exact}
\end{align}
where the flow time $t$ has mass dimension $-2$.
$I(t)$ is dimensionless, UV and IR finite ($p^2$ integral is well defined), and 
RG invariant.\footnote{
The OPE of a conceptually similar observable, 
$\langle \delta\alpha(x)^2 \rangle$ with a UV cutoff regularization,
has been studied in refs.~\cite{Bardeen:1976zh,David:1982qv,David:1985xj}.
The gradient flow allows us to
construct, in a simple and more sophisticated way, 
a UV and IR finite quantity which
has IR renormalons in its perturbative evaluation.
A similar observable in a nonlinear $\sigma$ model
has been considered in ref.~\cite{Ishikawa:2019tnw} in
the gradient-flow regularization, which discusses 
the renormalon structure of the $CP^N$ theory on $\mathbb{R}\times S^1$.
}
In this example, we give its dual transform by
\begin{align}
&
\tilde{I}(p^2)=\frac{1}{2\pi i}\int_{t_0-i\infty}^{t_0+i\infty}\!\!\!\!dt \,e^{tp^2}\frac{I(t)}{t^2}
=N\frac{D_\alpha(p^2)}{4\pi}   \label{Ip-InverseLaplace}
\,,
\\ &
I(t)
=t^2\int_0^\infty{dp^2}\, 
\tilde{I}(p^2)
\,e^{-tp^2}\,, \label{It-Laplace}
\end{align}
with the choice $(a,u')=(1,-2)$.  Recall Eq.~\eqref{eq:exactDT}.

The OPE of $\tilde{I}(p^2)$ (the expansion in $m^2/p^2$) is given by
\bea
&&
\tilde{I}(p^2)=p^2 \Biggl[\bar{g}(p^2)+\frac{m^2}{p^2} \lt( 2\bar{g}(p^2) -2 \bar{g}(p^2)^2 \rt) 
\nonumber\\&&
~~~~~~~~~~~~~~~~~
+\lt(\frac{m^2}{p^2} \rt)^2 \lt(-2 \bar{g}(p^2)-\bar{g}(p^2)^2+4 \bar{g}(p^2)^3 \rt)+\cdots \Biggr]
\,. 
\label{Ip-OPE}
\eea
IR renormalons are absent 
in all the Wilson coefficients in the dual space
because the perturbative series for the Wilson coefficients terminate at finite orders when expressed with $\bar{g}(p^2)$ \cite{Novikov:1984ac}.

In comparison, the OPE of $I(t)$ has renormalons.
The perturbative expansion in $\bar{g}(\mu^2)$ of the LO Wilson coefficient of
$I(t)$ is given by the Laplace transform 
of that of $\tilde{I}(p^2)$,
and its Borel transform is
expressed by\footnote{
Here, we define the Borel transform in the nonlinear $\sigma$
model by replacing
$\alfs(\mu^2)$ with $\bar{g}(\mu^2)$ and $b_0$ with $1$
in Eqs.~\eqref{eq:c0pt1} and \eqref{BorelTr-C0-3.3}.
}
\be
(\mu^2 t)^{u}\, \Gamma(2-u) .\label{It-PT-renormalon}
\ee
This shows that the IR renomalons are located at $u=2, 3, 4, \cdots$
and the leading renormalon uncertainty is $\mathcal{O}(t^2 m^4)$.
There are no UV renormalons due to the gradient-flow regularization.
These IR renormalons stem from the $p^2$-integral of $\log(\mu^2/p^2)^n$
encoded in $\bar{g}(p^2)$. 
They should eventually cancel since
$I(t)$ is defined unambiguously.

In the dual space approach, 
we apply a dual transform to an observable under consideration 
whose perturbative expansion has IR renormalons. 
We therefore suppose that $I(t)$ is an observable of our interest and 
we apply the dual transform as Eq.~\eqref{Ip-InverseLaplace} 
(although the above procedure may look reversed).
The parameters
$(a, u')=(1,-2)$ in Eq.~\eqref{Ip-InverseLaplace}
are chosen
to suppress the renormalons at $u=2, 3, 4, \cdots$.
It is notable that this is an example where all the Wilson coefficients in the dual-space OPE are free of renormalons,
which is the situation assumed in Sec.~\ref{ss:MB1}.
(This feature may seem obvious in this example, and we 
consider another example in Sec.~\ref{sec3:SEofSigma} to compensate for this drawback.) 


Since $D_{\alpha}(p^2)$ is known exactly even in the IR region, 
we can explicitly examine how renormalon cancellation takes place in the dual space approach and 
can check the validity of this framework in detail, as discussed in
this and the following subsections.

\subsubsection*{Expansion-by-regions and renormalon cancellation}

The OPE is obtained by EBR of the $p^2$-integral in Eq.~\eqref{It-Laplace}:
\begin{align}
&
\big[I(t)\big]_{\rm OPE}
=\lim_{\delta\to 0}
\left\{ I^{\pm}_h(t)+I^{\pm}_s(t)\right\}
\,,
\label{EBR-I(t)1}
\\
&
I^{\pm}_h(t)=t^{2-\delta}\int_{C_\mp}dp^2\, (p^2)^{-\delta_\pm}
\left[\tilde{I}(p^2)\right]_{\rm OPE}\,e^{-tp^2}
\,,
\label{EBR-I(t)2}
\\
&
I^{\pm}_s(t)=t^{2-\delta}\int_{C_\mp}dp^2\, (p^2)^{-\delta_\pm}
\tilde{I}(p^2)\,\left\{
1-tp^2+\frac{1}{2}(tp^2)^2-\dots
\right\}
\label{EBR-I(t)3}
\,,
\end{align}
where
\be
\delta_\pm=\delta\pm i0\,,
~~~~~
\delta\in\mathbb{R}\backslash\{ 0 \}
\,.
\ee
The EBR realizes the cancellation of
IR renormalons.
IR and UV divergences also appear, which are
regularized by $\delta$ and canceled in $\big[I(t)\big]_{\rm OPE}$.

For a while we focus on the first few terms of the OPE of $I(t)$:
\be
\big[I(t)\big]_{\rm OPE}
=C_0^{(h)}(t)+C_1^{(h)}(t) \langle\alpha\rangle t+C_{2}^{(h)}(t)\langle\alpha\rangle^2 t^2
+C_{2}^{(s)} \langle\delta\alpha^2\rangle t^2+{\cal O}(t^3) .
\label{It-OPE}
\ee
Explicitly they are given by
\begin{align}
C_0^{(h)}(t)&=t^2 \int_{C_{\mp}} d p^2 p^2 \bar{g}(p^2) e^{-t p^2}  ,  \\
C_1^{(h)}(t)&=t \int_{C_{\mp}} d p^2 \, \lt(2\bar{g}(p^2)-2 \bar{g}(p^2)^2 \rt) e^{-t p^2} , \\
C_2^{(h)}(t)&=t^{-\delta} \int_{C_{\mp}}  \frac{d p^2}{(p^2)^{1+\delta_{\pm}}} \,  \lt(-2 \bar{g}(p^2)- \bar{g}(p^2)^2+4\bar{g}(p^2)^3 \rt) e^{-t p^2} , \label{C2h}
\end{align}
and
\be
\langle \delta \alpha^2 \rangle
=\int_{C_{\mp}} dp^2 (p^2)^{-\delta_{\pm}} \tilde{I}(p^2)
\,,
\ee
with $C_2^{(s)}=1$. 
The Wilson coefficients are denoted as $C_i^{(h,s)}$ according to
their origins. We note that we applied EBR to the $p^2$-integral of the Laplace transform
and not to loop integrals associated with Feynman diagrams. That is why we have 
a soft--origin Wilson coefficient $C_2^{(s)}$. 
There appear two nonperturbative matrix elements,
$\langle\alpha(x)\rangle=m^2$ and 
$\langle\delta\alpha(x)^2\rangle={\cal O}(m^4)$, up to ${\cal O}(t^2)$.
We omit $\delta_{\pm}$ when not necessary (in the absence of IR/UV divergences).

Each term of the OPE \eqref{It-OPE} has ambiguities, which stem
from the regions where $p^2$-integrals have divergences.
The regions of $p^2$ which cause ambiguities are as follows:
\begin{align}
C_0^{(h)}:&~   p^2 \sim m^2 , \non
 C_1^{(h)}:&~ p^2 \sim m^2  , \non
  C_2^{(h)}:&~ p^2 \sim m^2 , \, p^2 \sim 0 ,  \non
 \langle \delta \alpha^2 \rangle:&~ p^2 \sim \infty .
\end{align}
The hard quantities have ambiguities associated with the Landau pole singularity ($p^2 \sim m^2$)
and IR divergences ($p^2 \sim 0$), whereas 
the soft quantities (or condensate) have UV divergences ($p^2 \sim \infty$).
In other words, the ambiguities originate from the region where their original  
expansions are {\it{not}} valid. 
By focusing on the contributions up to $\mathcal{O}(t^2 m^4)$ in the OPE,
these imaginary part and divergences are given by
\begin{align}
&C_0^{(h)}|_{\rm Landau}=t^2 \int_{p_1^2\mp i0}^{p_2^2 \mp i0} dp^2 p^2 \bar{g}(p^2)
=\pm i \pi t^2 m^4+\cdots \,,
\label{C0-Landau}
\\
&C_2^{(h)}|_{\rm Landau}= \int_{p_1^2\mp i0}^{p_2^2 \mp i0} 
 \frac{d p^2}{p^2} \,  \lt(-2 \bar{g}(p^2) \rt)
 =\mp 2 i \pi   +\cdots \,,
\\
&C_2^{(h)}|_{\rm soft}
=t^{-\delta} \int_0^{p_1^2}
 \frac{d p^2}{(p^2)^{1+\delta_{\pm}}} \,  \lt(-2 \bar{g}(p^2) \rt) 
=2  \int_0^{\infty} d\alpha \lt( \frac{p_1^2}{m^2} \rt)^{\alpha}  \frac{1}{\alpha-\delta_{\pm}} \non
&~~~~~~~~~~=-2  \log{(-\delta_{\pm}})+\cdots \,,
 \label{C2-soft}
\end{align}
\begin{align}
\langle \delta \alpha^2 \rangle|_{\rm hard} 
&=\int_{p_2^2}^{\infty} d p^2 (p^2)^{-\delta_{\pm}}\lt[ p^2 \bar{g}(p^2) +\frac{m^4}{p^2} (-2 \bar{g}(p^2)) \rt] \non
&=\mp i \pi m^4+2 m^4 \log{\delta_{\pm}}+\cdots \,,
 \label{cond-hard}
\end{align}
where $0< p_1^2 < m^2 < p_2^2  <1/t$.
In the above equations, the dots
``$\cdots$'' represent real finite contributions.
In evaluating the integrals for $C_{0,2}^{(h)}$ around the Landau pole we expand the exponential factor $e^{-t p^2} \simeq 1$.
The ambiguities found in $C_{0}^{(h)}|_{\rm Landau}$ are
the same as the renormalon uncertainties 
derived from Eq.~\eqref{It-PT-renormalon}.
It turns out that $C_1^{(h)}|_{\rm Landau}$ has higher order imaginary part 
which we neglect here.
Also the higher order terms in $\bar{g}(p^2)$ in the integrand for 
$C_2^{(h)}|_{\rm Landau}$ give higher order imaginary part.
We note that, in evaluating the integral for $\langle \delta \alpha^2 \rangle$
in the region $p^2\sim\infty$, it is justified to expand 
$\tilde{I}(p^2)$ in $m^2/p^2$ as in Eq.~\eqref{Ip-OPE}.
We therefore have the same integrands in the condensate \eqref{cond-hard} 
as those in the Wilson coefficients \eqref{C0-Landau} and \eqref{C2-soft}. 
In the integrand of Eq.~\eqref{cond-hard}, we only show the terms which give imaginary part or UV divergences.
The imaginary part in Eq.~\eqref{cond-hard} arises as \cite{Beneke:1998ui}
\begin{align}
\int_{p_2^2}^{\infty} d p^2 (p^2)^{1-\delta_{\pm}} \bar{g}(p^2)
&=\int_0^{\infty}  d \alpha \,  m^{2 \alpha} \int_{p_2^2}^{\infty} d p^2 (p^2)^{1-\delta_{\pm}-\alpha} \non
&=\int_0^{\infty}  d \alpha \, (p_2^2)^{2-\alpha-\delta_{\pm}} m^{2 \alpha} \frac{1}{\alpha+\delta_{\pm}-2}
=\mp i \pi m^4 +\cdots \,,
\end{align}
and the appearance of the $\log{\delta_{\pm}}$ can be understood
by a similar calculation to Eq.~\eqref{C2-soft}.
One can see that the above imaginary part and
$\log{\delta}$ terms found in Eqs.~\eqref{C0-Landau}--\eqref{cond-hard} cancel in the OPE \eqref{It-OPE}, as expected.

The above cancellation indicates the validity of the EBR method.
In fact the validity of the EBR relation Eqs.~\eqref{EBR-I(t)1}--\eqref{EBR-I(t)3}
can be proved following the strategy
of ref.~\cite{Jantzen:2011nz}.  
(See App.~\ref{App-ProofEBR}.)
The crucial identity which guarantees the validity is given by\footnote{
This identity can be shown by analytic continuation of $a_\pm$:
Change the integral variable 
to $z=\log(p^2/m^2)$, and close the
contour in the upper/lower half $z$-plane
for $a_R= -1$, $a_I>0$.
Note that it is crucial to take the sign of $C_{\pm}$ and that of 
$a_{\pm}$ to be the same.
}
\be
\int_{C_\pm} dp^2 \, \frac{(p^2)^{a_\pm}}{\{\log(p^2/m^2)\}^n}
=0
\,,
\label{FundamentalId-EBR}
\ee
where 
\be
n\in\mathbb{Z}_{>0}\,,
~~~~
a_\pm\equiv a_R\pm ia_I~~\text{with}~~a_R,a_I\in\mathbb{R},
~~a_I>0
\,.
\ee
The cancellation explicitly confirmed above
is a consequence of
the particular identities 
\be
\int_{C_{\mp}} dp^2 (p^2)^{-\delta_{\pm}} p^2 \bar{g}(p^2)=0 ,
\ee
\be
\int_{C_{\mp}} dp^2 (p^2)^{-1-\delta_{\pm}} (-2 \bar{g}(p^2))=0 ,
\ee
which lead to $\pm i \pi m^4 \mp i \pi m^4=0$ and
$\mp 2 i \pi -2 \log{(-\delta_{\pm})}+2\log{\delta_{\pm}}=0$, respectively.
Eq.~\eqref{FundamentalId-EBR} also explains the simultaneous absence of
the imaginary part associated with the Landau pole $C_1^{(h)}|_{\rm Landau}
=\int_{p_1^2 \mp i0}^{p_2^2 \mp i0} d p^2 (p^2)^{-\delta_{\pm}} \, \lt(2\bar{g}(p^2)-2 \bar{g}(p^2)^2 \rt) $
and the UV divergence in the condensate, 
which could be caused by 
$\int_{p_2^2}^{\infty} d p^2 (p^2)^{-\delta_{\pm}} \, \lt(2\bar{g}(p^2)-2 \bar{g}(p^2)^2 \rt) $;
if either of them is zero, the other should also be zero.

\begin{figure}[tbp]
\centering
\includegraphics[width=6cm]{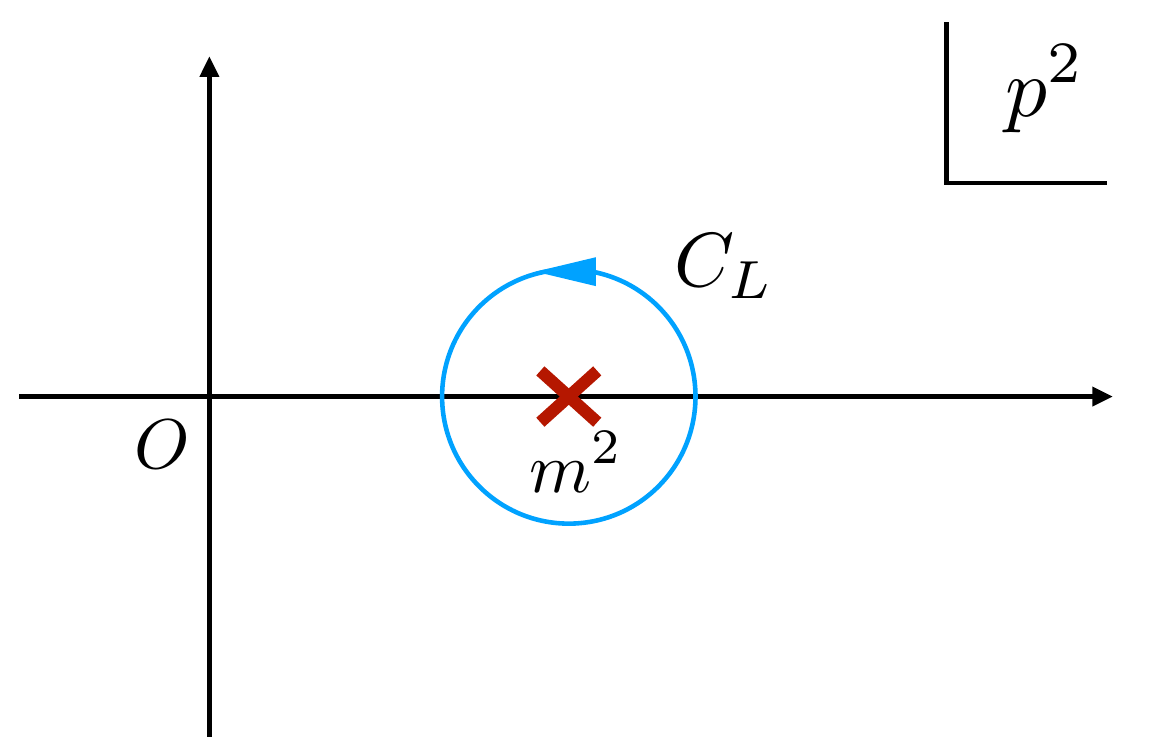}
\caption{\small
Contour $C_L$ surrounding the Landau pole 
in the complex $p^2$ plane.
}
\label{Fig-CL}
\end{figure}

Using the relation \eqref{FundamentalId-EBR},
we can discuss in a general manner the cancellation of the imaginary part
in $I_h^\pm(t)$ and $I_s^\pm(t)$ which
are related to renormalons.
As seen explicitly in the first few terms of the OPE, the imaginary part arise from three regions:
(i) $p^2\to 0$, (ii) $p^2\sim m^2$, (iii) $p^2\to\infty$.
The imaginary part from  (i) and (ii) are included in 
the Wilson coefficients of $I_h^\pm$, while that from
(iii) is included in the matrix elements of $I_s^\pm$.\footnote{
These features are somewhat confusing, since they stem from ``IR region of the hard
contribution'' and ``UV region of the soft contribution,'' respectively.
}
In fact, the above relation ensures the following identity:
\be
\bigg(\int_0^{p_1^2}+\int_{p_2^2}^\infty\bigg) dp^2\,\bigg(\frac{m^2}{p^2}\bigg)^n\frac{(p^2)^{1+\ell-\delta_\pm}}{\log(p^2/m^2_\pm)^{k+1}}
=-\int_{p_1^2}^{p_2^2}dp^2\,\bigg(\frac{m^2}{p^2}\bigg)^n\frac{(p^2)^{1+\ell-\delta_\pm}}{\log(p^2/m^2_\pm)^{k+1}},
\label{ren-Landau}
\ee
where $n,k,\ell\in\mathbb{Z}_{\ge 0}$. 
$n$ and $k$ correspond to the expansion orders of $\tilde{I}(p^2)$ in $m^2/p^2$ and $\bar{g}(p^2)$,
respectively, i.e., the ${\cal O}(\bar{g}(p^2)^{k+1}/p^{2(n-1)})$ term of $\tilde{I}(p^2)$,
whereas $\ell$ is an expansion order of $e^{- t p^2}$ in $t p^2$.
Note that in the regions where imaginary part or
IR/UV divergences arise it is justified to expand
$\tilde{I}(p^2)$ in $m^2/p^2$ and $\bar{g}(p^2)$, and $e^{- t p^2}$ in $t p^2$;
for this reason it is sufficient to assume the form of the integrand
of Eq.~\eqref{ren-Landau}.
The integrals on both sides are 
the contributions to ${\cal O}(t^{\ell+2})$ term of $I(t)$.
For notational convenience, we have taken the integral path on the
positive real axis and instead shifted the Landau pole infinitesimally
above or below the axis, $m^2_\pm\equiv m^2\pm i0$.
The imaginary part of the LHS of Eq.~\eqref{ren-Landau} is calculated as
\begin{align}
&{\rm Im}\int_0^\infty \!\!\!d\alpha\bigg[\int_0^{p_1^2}\!\!\!dp^2\frac{(-\alpha)^k(p^2)^{1+\ell-\delta_\pm}}{k!}\bigg(\frac{m^2_\pm}{p^2}\bigg)^{n-\alpha}\!
+\int_{p_2^2}^\infty \!\!\!dp^2\frac{\alpha^k(p^2)^{1+\ell-\delta_\pm}}{k!}\bigg(\frac{m^2_\pm}{p^2}\bigg)^{n+\alpha}\bigg]\non
&=\pm \pi (m^2)^{\ell+2}\frac{(\ell+2-n)^k}{k!}\big(\delta_{\ell+2-n\leq0}+\delta_{\ell+2-n>0}\big).
\label{SumKroneckerDelta}
\end{align}
The first term is from the hard contribution (in $I_h^\pm$)
and the second is from the soft contribution (in $I_s^\pm$).\footnote{
In the last line of Eq.~\eqref{SumKroneckerDelta},
we chose $\delta>0$.
The sum is unchanged if we choose $\delta<0$.
}
Note that each contribution is independent of $p_1^2,p_2^2$,
as it should be.
The RHS of Eq.~\eqref{ren-Landau} contains the imaginary part due to 
the Landau pole of the running coupling constant
$\bar{g}(p^2)$, which is calculated as
\begin{align}
&\mp\frac{1}{2i}\bigg(\int_{p_1^2+i0}^{p_2^2+i0}-\int_{p_1^2-i0}^{p_2^2-i0}\bigg)dp^2\,\bigg(\frac{m^2}{p^2}\bigg)^n\frac{(p^2)^{1+\ell-\delta_\pm}}{\log(p^2/m^2)^{k+1}}\non
&=\pm\frac{1}{2i}\int_{C_L}dp^2\,\bigg(\frac{m^2}{p^2}\bigg)^n\frac{(p^2)^{1+\ell-\delta_\pm}}{\log(p^2/m^2)^{k+1}}
=\pm \pi (m^2)^{\ell+2}\frac{(\ell+2-n)^k}{k!}
\,,
\label{contr-region(ii)}
\end{align}
where $C_L$ is the closed path surrounding the Landau pole 
counterclockwise in the complex $p^2$ plane (Fig.~\ref{Fig-CL}).
These imaginary part from different regions appear to be unrelated at a first glance,
but the EBR relation shows that there is a one-to-one correspondence 
among them at each order term of the OPE and coupling expansion,
namely before summing over $n,k,\ell$.
Using this relation, we can show that the imaginary part of 
$I_h^\pm(t)+I_s^\pm(t)$ is identically zero,
consistently with the fact that $I(t)$ is real:
\bea
&&{\rm Im}\bigg[I_h^\pm(t)+I_s^\pm(t)\bigg]\non
&&=\pm\pi m^4t^2\sum_{\ell=0}^\infty\frac{(-m^2t)^\ell}{\ell!}\sum_{n=0}^\infty \sum_{k=0}^n\frac{d_{n,k}(0)}{k!}(\ell+2-n)^k\big(\delta_{\ell+2-n\leq0}+\delta_{\ell+2-n>0}\big)\non
&&~~~\mp\pi m^4t^2\sum_{\ell=0}^\infty\frac{(-m^2t)^\ell}{\ell!}\sum_{n=0}^\infty \sum_{k=0}^n\frac{d_{n,k}(0)}{k!}(\ell+2-n)^k
\non
&&=0
\,,
\eea
where $d_{n,k}(0)$ represent perturbative coefficients appearing in $\tilde{I}(p^2)$.
See Eq.~\eqref{def-Xn-dnk(0)}.
The upper bound of the summation in $k$ is equal to $n$ because 
$d_{n,k}(0)=0$ for $k>n$.
The above conclusion does not depend on these coefficients.
This shows how the imaginary part cancel in the entire OPE
of $I(t)$.
Although different reasonings justify the
expansion in $n,k,\ell$ in the individual regions in the $p^2$ space
(e.g., to extract the UV behavior of the soft expansion), essentially the
factorization scale independence
between the neighboring regions 
underlies the cancellation 
mechanism.

There is another nontrivial relation regarding the imaginary part.
When summed over $n$ and $k$, the imaginary part due to the 
 Landau pole vanishes at each order of $t$.
Namely, the coefficient of $(m^2t)^\ell$ vanishes as
\begin{align}
&\sum_{n=0}^\infty \sum_{k=0}^n\frac{d_{n,k}(0)}{k!}(\ell+2-n)^k
=
\sum_{n=0}^\infty 
X_n(\ell+2-n)
\non
&
=\sum_{n=0}^{2\ell+4}(-1)^n
\frac{2(6+2n^2+7\ell+2\ell^2-4n(2+\ell))}
{(4+2\ell)(3+2\ell)}\,_{4+2\ell} C_n
=0
\,,
\label{ren-can-cond}
\end{align}
for arbitrary $\ell\in \mathbb{Z}_{\ge 0}$.
We used the explicit result for $d_{n,k}$ given below [Eq.~\eqref{Borel-tildeCn}].
The last equality is shown using
an identity $\sum_{n=0}^N {_N} C_n(-1)^n \, n^j =0$
for $j<N$.
This feature can be understood as follows.
If we take the contour $C_L$ 
sufficiently away from the Landau pole
in the right half plane, the integrand becomes a convergent
series everywhere
along the contour $C_L$ in Eq.~\eqref{contr-region(ii)}.
Since 
before the OPE $\tilde{I}(p^2)$ is regular and has no Landau pole
inside the contour, 
the integral (residue) should converge to zero
by summation over $n$ and $k$.
It is interesting to note that due to the identity \eqref{contr-region(ii)} 
this argument shows that also the sum of the imaginary part from the
other two regions vanishes.
Therefore the absence of the Landau pole in $\tilde{I}(p^2)$
guarantees the fact that $I(t)$ is real.

We make a supplementary comment on the cancellation mechanism of the Landau pole singularity.
The cancellation of the Landau pole, which is a physical requirement, is realized not within perturbation theory (the expansion in $\bar{g}(p^2)$) but within the double expansion in $\bar{g}(p^2)$ and $m^2/p^2$ in the dual-space. The clear observation of the cancellation is allowed by the feature that the double expansion, in particular the perturbative expansion, is well defined in the dual space due to the absence of renormalons. 
If the perturbative expansion were asymptotic, 
as in the original $t$ space, it would be difficult to examine the cancellation 
of the Landau pole in the double expansion.

We showed that the cancellation of $\log{\delta}$ terms and of imaginary part 
take place in this example at any order of the OPE.
This understanding leads us to a scheme appropriate for DSRS
of the Wilson coefficients:
(1)~Obtain a converging series in the dual space, corresponding to the dual-space
OPE. 
(2)~Apply the Laplace transform to the dual-space OPE, which provides 
the Wilson
coefficients of the hard contribution. 
(3)~In the Laplace transform, IR divergences are regularized by 
$p^{-2\delta_\pm}$ and the $\log\delta$ terms are
subtracted minimally by counter terms. 
(4)~Adapt the PV
prescription, i.e., remove the imaginary part. 
(5)~The nonperturbative matrix elements are
treated as parameters. 
(If necessary, one can resum UV renormalons in addition, see 
Sec.~\ref{sec3:SEofSigma}.)
This DSRS scheme naturally encodes the cancellation between 
IR and UV divergences and the renormalon cancellation.
(It adds renormalization of IR divergences to
the conventional PV prescription.)


\subsubsection*{Perturbative coefficients to all orders in the OPE}

Below we present explicit formulas of the perturbative coefficients
to all orders in the OPE. 
(See App.~\ref{App-Calc-NLsigma} for details of the calculations.)
So far, we have treated the $p^2$-integral with the integrands
expressed with the running coupling constant $\bar{g}(p^2)$.
Here, we mainly study expressions with the 
fixed-scale coupling constant $\bar{g}(\mu^2)$.
The OPE of $\tilde{I}(p^2)$ is expressed in the form
\begin{align}
&
\left[\tilde{I}(p^2)\right]_{\rm OPE}=
p^2\sum_{n=0}^\infty \biggl(\frac{m^2}{p^2}\biggr)^n\,\tilde{C}_n(p^2)
\,,
\label{tildeI_OPE}
\end{align}
with the Wilson coefficients in the dual space given by
\begin{align}
\tilde{C}_n(p^2)=\sum_{k=0}^{\infty}\bar{g}(\mu^2)^{k+1}
\, d_{n,k}(L_p)\,,
~~~~~
L_p=\log\left(\frac{\mu^2}{p^2}\right)
\,.
\label{WilsonCoeff-tildeI}
\end{align}
They are obtained by expanding Eq.~\eqref{Dalpha}
in $m^2/p^2$ and $1/\log(p^2/m^2)=\bar{g}(p^2)$,
and then by re-expressing $\bar{g}(p^2)=\bar{g}(\mu^2)/[1+\bar{g}(\mu^2)\log(p^2/\mu^2)]$ in
a geometric series of $\bar{g}(\mu^2)$.
Inserting Eq.~\eqref{tildeI_OPE} into Eq.~\eqref{EBR-I(t)2}
we obtain the Wilson coefficients of the hard contribution,
\be
I^{\pm}_h=\sum_{n=0}^\infty (m^2t)^n
\left[C_{n}^{(h)}(t)\right]_\pm
\,,
~~~~~
\left[C_{n}^{(h)}(t)\right]_\pm
\!=
t^{2-\delta-n}\!\int_{C_\mp}\!\!\!
dp^2\, (p^2)^{1-\delta_\pm-n}
\tilde{C}_n(p^2)\,e^{-tp^2}
\,.
\ee

The Borel transform of $[\tilde{I}(p^2)]_{\rm OPE}$ is given by
\be
B_{\tilde{I}}(u)=
\sqrt{p^2(p^2+4m^2)}
\left(\frac{\sqrt{p^2+4m^2}+\sqrt{p^2}}{2\mu}\right)^{-2u}
=
p^2\sum_{n=0}^\infty \biggl(\frac{m^2}{p^2}\biggr)^n\,B_{\tilde{C}_n}(u)
\,.
\ee
The Borel transform of each Wilson coefficient is given by
\begin{align}
B_{\tilde{C}_n}(u)
&=\sum_{k=0}^\infty\frac{d_{n,k}(L_p)}{k!}u^k=
\biggl(\frac{\mu^2}{p^2}\biggr)^u
\frac{(4u^2-2u-2n)\Gamma(-n-2u)}{\Gamma(n+1)\Gamma(2-2u-2n)}
\non
&=
\biggl(\frac{\mu^2}{p^2}\biggr)^u X_n(u)
\,,
\label{Borel-tildeCn}
\end{align}
where we define an $n$-th degree polynomial of $u$ as
\begin{align}
X_n(u)&=\sum_{k=0}^n\frac{d_{n,k}(0)}{k!}u^k
=\frac{2(-1)^n}{n!}(2u^2-u-n)(1+n+2u)_{n-2}
\,.
\label{def-Xn-dnk(0)}
\end{align}
$(a)_j=\Gamma(a+j)/\Gamma(a)$ denotes the Pochhammer symbol.
We confirm that $B_{\tilde{I}}(u)$ has no singularities on the positive
real $u$ axis. (In fact, it has no singularities in the entire
$u$ plane.)
Hence,
$[\tilde{I}(p^2)]_{\rm OPE}$
does not have IR renormalons,
even though it is rather obvious from the definition of $I(t)$.
We can also calculate $d_{n,k}(L_p)$ from Eq.~\eqref{Borel-tildeCn}.

The Borel transform of $I_h^\pm(t)$ is given by
\be
B_{I_h}(u)=
\sum_{n=0}^\infty (m^2t)^n\, B_{C_n^{(h)}}(u)
\,,
\ee
where the Borel transform of each
Wilson coefficient is expressed in terms
of $X_n(u)$ as
\be
B_{C_n^{(h)}}(u)=
(\mu^2t)^u\,
\Gamma(2-n-u-\delta)\,
X_n(u)
\,.
\label{B_Cnh}
\ee
We find that
$B_{C_0^{(h)}}$
has poles at $u=2,3,4,\cdots$;
$B_{C_1^{(h)}}$
at $u=2,3,4,\cdots$;
$B_{C_2^{(h)}}$
 at $u=1-\delta,2-\delta,3-\delta,\cdots$,
and so on.
(For $n<2$ we can safely set $\delta$ to zero.)
Thus, $I^\pm_h(t)$ has IR renormalons.
In addition, expanding $B_{C_n^{(h)}}$ in $u$, we see that the
perturbative expansion of $C_n^{(h)}$ in $\bar{g}(\mu^2)$
for $n\ge 2$ includes IR divergences
as multiple poles in $\delta$.
This originates from the $u\sim 0$ region,
by expansion of $B_{C_n^{(h)}}\sim 1/(u+\delta)$ in $u$.

$I^\pm_h(t)$ is constructed by Borel resummation as
\be
I^\pm_h(t)=\int_{C_\pm}du\,B_{I_h}(u)\,e^{-u/\bar{g}(\mu^2)}
\,.
\ee
As already stated, this is equivalent to Eq.~\eqref{EBR-I(t)2}.
For example, we obtain the LO Wilson coefficient as
\be
\left[C_{0}^{(h)}(t)\right]_\pm=\int_{C_\pm} du\,(\mu^2t)^u
\,\Gamma(2-u)\,e^{-u/\bar{g}(\mu^2)}
\,,
\ee
from Eq.~\eqref{B_Cnh} 
for $n=0$, as given by eq.~\eqref{It-PT-renormalon}.
We can insert $1/\bar{g}(\mu^2)=\log(\mu^2/m^2)$, express 
$\Gamma(2-u)$ as $\int_{C_\mp} dx \, x^{1-u}\,e^{-x}$,
 transform the integral variable from $x$ to $p^2=x/t$,
and integrate over $u$.
In this way we derive the dual space expression
\be
\left[C_{0}^{(h)}(t)\right]_\pm
=t^2\int_{C_\mp}dp^2\,\frac{p^2\,e^{-tp^2}}{\log(p^2/m^2)}
\,.
\ee

Although IR divergence appears in $\bigl[C_{n}^{(h)}\bigr]_\pm$
for $n\ge 2$ as multiple poles in $\delta$ at fixed order as mentioned above,
by Borel resummation
they turn to $\sim \log\delta$.
In the Borel integral, this originates from the $u\sim 0$
region, where the integrand behaves as $\sim 1/(u+\delta)$.
In the dual space integral, the same $\log\delta$
divergence stems from the
$p^2\sim 0$ region.

The soft contribution can be expressed as
\be
I_s^\pm(t)=\sum_{\ell=0}^\infty \frac{t^{\ell-\delta+2}}{\ell!}\,
\langle \delta\alpha\, \Box^{\ell-\delta_\pm}\delta\alpha \rangle_\pm
\,,
\ee
by Eq.~\eqref{EBR-I(t)3} and the definition of $D_\alpha(p^2)$.
All the matrix elements have UV divergences,
$\langle \delta\alpha\, \Box^{\ell-\delta_\pm}\delta\alpha \rangle_\pm
\sim \log\delta$, which stem 
from the region $p^2\to\infty$
and cancel the IR divergences in $I_h^\pm$.
Thus, 
\be
I_s^\pm(t) = \sum_{\ell=0}^\infty C^{(s)}_{\ell+2} \, t^{\ell+2} \, 
\langle \delta\alpha\, \Box^{\ell-\delta_\pm}\delta\alpha \rangle_\pm
\,,
~~~~~
C^{(s)}_{\ell+2} =\frac{1}{\ell !} \,.
\ee
The Wilson coefficients of the soft contribution are
$t$ independent and finite as $\delta\to 0$.

Thus, the Wilson coefficients of the hard contribution,
$[C_n^{(h)}]_\pm$, have imaginary part
whose sign depends on the regularization.
In addition $[C_n^{(h)}]_\pm$ for $n\ge2$ contains an IR divergence which behaves as
$\log\delta$.
This is canceled by the UV divergence contained in
the matrix element $\langle\delta\alpha\,\Box^{n-2-\delta_\pm}\delta\alpha
\rangle_\pm$.
The cancellation of $\log\delta$ takes place in the
same order of $t^n$.
On the other hand, the cancellation of the imaginary part
is ensured at the level of the entire OPE, since renormalons
shift the order counting of $t$.

\subsection{DSRS method and fit of nonperturbative parameters (Simulation)}
\label{sec3:simulation-study}

Using the OPE of $I(t)$ obtained above,
let us simulate
practical application of the DSRS method to QCD.
In that case, 
we can only calculate the Wilson coefficients of 
$[I(t)]_{\rm OPE}$ using a finite number of perturbative coefficients. 
The nonperturbative matrix elements
need to be 
determined by comparison of 
$[I(t)]_{\rm OPE}$ with 
experimental data or with nonperturbative calculations by lattice QCD.
As we showed in the previous section, the EBR method
guarantees that the IR divergences and imaginary part in 
$[C_n^{(h)}]_\pm$ are canceled by the matrix elements.
Therefore, it is possible to determine the well-defined nonperturbative matrix 
elements in a
systematic  approximation 
by subtracting the imaginary part and the IR divergences
from the computation of the Wilson coefficients.

Specifically, 
we attempt to determine the nonperturbative parameters of
the OPE by comparing the (simulated)
experimental values with theoretical calculation of $I_h(t)$ which is defined by subtracting the IR divergence and imaginary part 
from $I_h^\pm(t)$ (Eq.~\eqref{EBR-I(t)2}).
The experimental value here is the exact $I(t)$ defined by Eqs.~\eqref{Dalpha} and \eqref{It-exact}, and there are no statistical uncertainties in this analysis.
Only systematic uncertainties due to truncation of $t$ expansion
and perturbative evaluation of the Wilson coefficients are involved and studied.
We perform an analysis of the OPE up to ${\cal O}(t^2)$, and
$I_h(t)$ is given by
\be
I_h(t) =\big[C^{(h)}_0(t)\big]_{\rm PV}+\big[C^{(h)}_1(t)\big]_{\rm PV}\langle\alpha\rangle t+\big[C^{(h)}_2(t)\big]_{\rm PV}^{\rm ren}\,\langle\alpha\rangle^2 t^2+{\cal O}(t^3),
\label{It-hard-t2}
\ee
\be
\big[C^{(h)}_0(t)\big]_{\rm PV}=t^2\int_{0,\rm PV}^\infty \!\!\!\!dp^2 \,p^2\sum_{k=0}^\infty d_{0,k}(L_p)\,\bar{g}(\mu^2)^{k+1}e^{-tp^2},
\label{simint1}
\ee
\be
\big[C^{(h)}_1(t)\big]_{\rm PV}=t\int_{0,\rm PV}^\infty \!\!\!\!dp^2 \,\sum_{k=0}^\infty d_{1,k}(L_p)\,\bar{g}(\mu^2)^{k+1}e^{-tp^2},
\ee
\be
\big[C^{(h)}_2(t)\big]_{\rm PV}^{\rm ren}=G^{\rm IR}(t,\Lambda_{\rm IR})+\int_{\Lambda_{\rm IR}^2,\rm PV}^\infty \!\!\frac{dp^2}{p^2} \,\sum_{k=0}^\infty d_{2,k}(L_p)\,\bar{g}(\mu^2)^{k+1}e^{-tp^2},
\label{simint3}
\ee
\bea
G^{\rm IR}(t,\Lambda_{\rm IR})
&=&\bigg[\int_0^{\Lambda_{\rm IR}^2} \!\!\frac{dp^2}{p^{2+2\delta_\pm}}\,\big[-2{\bar{g}(p^2)}-{\bar{g}(p^{2})}^2+4{\bar{g}(p^2)}^3\big]+2\log(-\delta_\pm)\bigg]_{\delta_\pm\to0}\non
&=&-2\gamma_E-\frac{2}{L_{\rm IR}^2}-\frac{1}{L_{\rm IR}}-2\log(L_{\rm IR}),
\quad\quad
L_{\rm IR}=\log(m^2/\Lambda_{\rm IR}^2).
\eea
In the above formulas, we have subtracted
the imaginary part and IR divergence from each 
Wilson coefficient 
by taking the principal value of the integral
and calculating the explicit form of the IR divergence.
The difference from $I_h^\pm(t)$ is given analytically by
\be
I_h^\pm(t)-I_h(t)=-(2\log\delta\,\pm i\pi)m^4t^2+{\cal O}(t^3)
\,,
\label{dif-Hpm}
\ee
which is RG invariant.
The Wilson coefficients in the above formulas are expressed in 
expansion in $\bar{g}(\mu^2)$.
In the analysis below, we will set $\mu^2\propto t^{-1}$ or $\mu^2\propto p^2$;
the coupling expansion of the integrand will be truncated 
at a finite order and the truncation error will be estimated.
In order to calculate the IR divergence explicitly in terms of $\delta$, 
$\big[C^{(h)}_2(t)\big]_{\rm PV}^{\rm ren}$ is separated into two contributions by introducing a cutoff $\Lambda_{\rm IR}$, although the cutoff dependence cancels out between them.
The IR divergence behaves as $\log\delta$, which is quite different from the typical
form of IR divergences.
The finite part that remains after subtracting the IR divergence is RG invariant.

\begin{figure}[tbp]
\centering
\includegraphics[width=13cm]{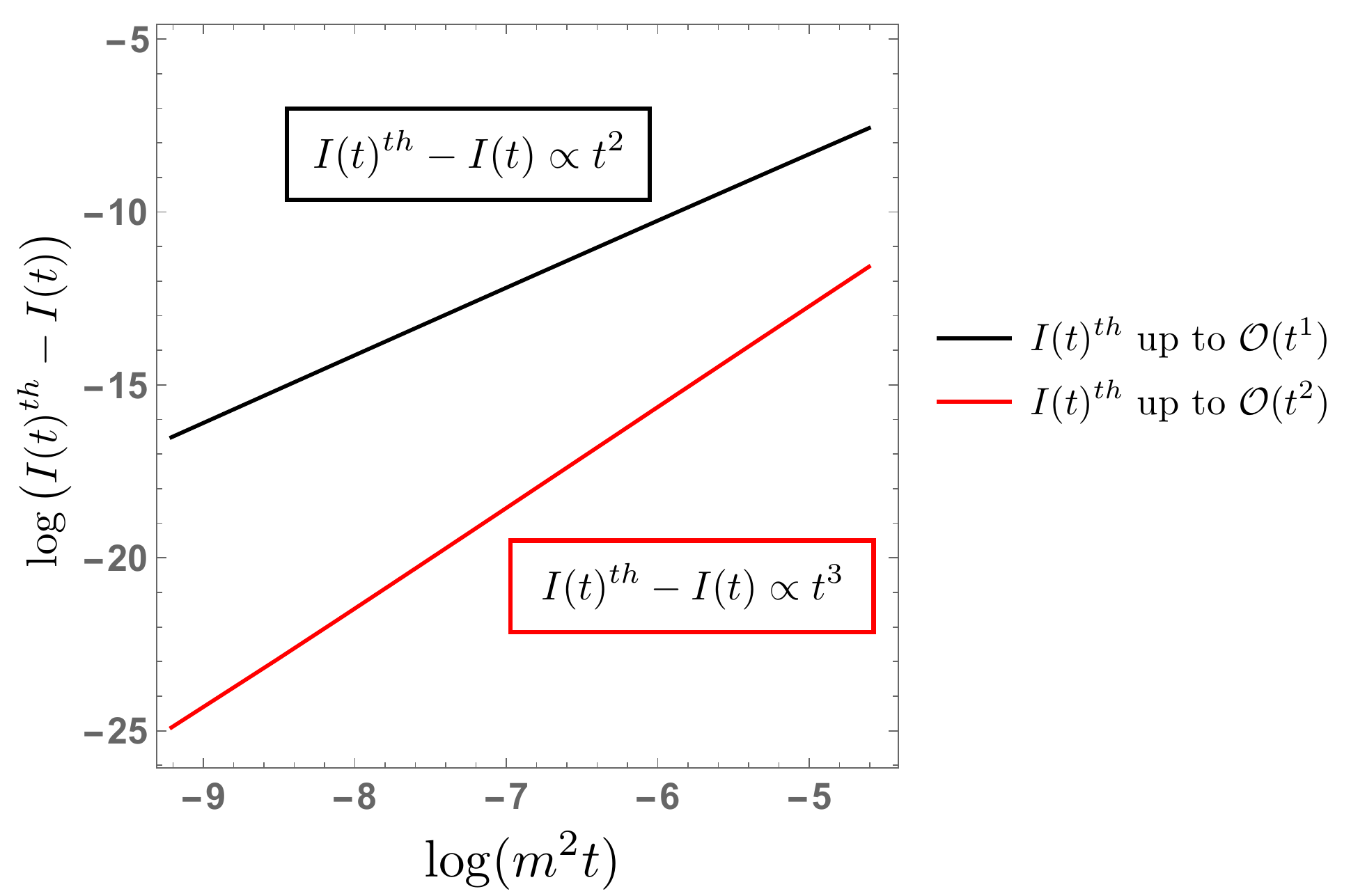}
\caption{\small
Log-log plot of
the difference between $I(t)$ and $I(t)^{th}=I_h(t)+\langle\delta\alpha^2\rangle_{\rm PV}^{\rm ren} t^2$.
The black (red) line corresponds to
$I(t)^{th}$ computed up to ${\cal O}(t^1)$ (${\cal O}(t^2)$). 
The slope of each line represents the power behavior of $t$.
}
\label{Ith-Iex}
\end{figure}

The soft contribution $I_s(t)$ can also be computed in this model.
Its expression up to ${\cal O}(t^2)$ is obtained as
\be
I_s(t)^\pm=\langle\delta\alpha^2\rangle_{\rm PV}^{\rm ren}\,
 t^2+(2\log\delta\pm i\pi)\,m^4 t^2+{\cal O}(t^3),
\label{It-soft-t2}
\ee
\vspace{-5mm}
\be
\langle\delta\alpha^2\rangle_{\rm PV}^{\rm ren}=\int_0^{\Lambda_{\rm UV}^2}\!\!\!\!dp^2\tilde{I}(p^2)+G^{\rm UV}(\Lambda_{\rm UV}),
\label{del-alf2}
\ee
\vspace{-5mm}
\bea
G^{\rm UV}(\Lambda_{\rm UV})&=&\bigg[\int_{\Lambda_{\rm UV}^2}^\infty \!\!{dp^2}{p^{2-2\delta_\pm}} \,\sum_{n=0}^2(m^2/p^2)^n\sum_{k=0}^\infty d_{n,k}(0)
\bar{g}(p^2)^{k+1}-2\log(\delta_\pm)\bigg]_{\delta\to0}\non
&=&2\gamma_E-{\rm Ei}\big(2 L_{\rm UV}\big) + \frac{2-L_{\rm UV}\big(1+2 (\Lambda_{\rm UV}^2/m^2)\big)}{L_{\rm UV}^2}+ 2\log\big(L_{\rm UV}\big),
\eea
where ${\rm Ei}(z)=-\int_{-z}^\infty dt\,e^{-t}/t$,
and
\be
L_{\rm UV}=\log(\Lambda_{\rm UV}^2/m^2).
\ee
To explicitly compute the UV divergence, we introduce a cutoff $\Lambda_{\rm UV}$, 
although 
$\langle\delta\alpha^2\rangle_{\rm PV}^{\rm ren}$ also has no overall cutoff dependence.
Comparing Eqs.~\eqref{dif-Hpm} and \eqref{It-soft-t2}, we can see that the IR and UV divergences and the imaginary part cancel explicitly.
$\langle\delta\alpha^2\rangle_{\rm PV}^{\rm ren}$ can be evaluated by numerical integration to be
\be
\frac{\langle\delta\alpha^2\rangle_{\rm PV}^{\rm ren}}{m^4}=-1.386\cdots\,.
\label{del-alf2-exact}
\ee

As a cross check,
we compare the exact (experimental) value of $I(t)$ and $I(t)^{th}=I_h(t)+\langle\delta\alpha^2\rangle_{\rm PV}^{\rm ren} \,t^2$
using $\langle\alpha\rangle=m^2$ and Eq.~\eqref{del-alf2-exact}.
The difference $I(t)^{th}-I(t)$ is shown in Fig.~\ref{Ith-Iex}.
We set $\mu^2=p^2$ in the integrand (in the dual space) and calculate
$I(t)^{th}$ up to ${\cal O}(t^1)$ and ${\cal O}(t^2)$, 
which correspond to the black and red lines, respectively.
The typical size of the coupling constant in this figure is $\bar{g}(1/\sqrt{t})\in[0.1,\,0.2]$.
We see that the exact value of $I(t)$ can be predicted systematically,
by increasing the expansion order of $t$ for
$I_h^\pm(t)$ and $I_s^\pm(t)$ and by the DSRS method.

We examine whether we can 
determine the nonperturbative parameters of the OPE correctly
by a fit to an experimental data set.
Here, we choose the position of the 
Landau pole, 
$\langle\alpha\rangle=m^2$, and $\langle\delta\alpha^2\rangle_{\rm PV}^{\rm ren}$ 
given by Eq.~\eqref{del-alf2}, as the fit parameters.
We consider the case where the coupling constant (or equivalently, the position
of the Landau pole) needs to be determined by the fit as well.
We compare a finite number of data points
from theoretical computation of $I_h(t)$ and from exact (experimental) values of $I(t)$.
The experimental data is given by
\be
I_{\rm exp}(t_{\rm exp};r)=t_{\rm exp}^2\int_0^\infty dp^2\tilde{I}(p^2)\,e^{-t_{\rm exp}p^2},
\quad\quad
t_{\rm exp}=r^2\,t.
\ee
$r$ defines a scale unit, which is equal to the ratio of $m$ and
the position of the Landau pole.
(In this case $r=1$.)
On the other hand,
$I_h(t)$ is obtained using the Wilson coefficients computed from 
a finite number of perturbative coefficients.
We calculate the Wilson coefficients in two different ways, 
in order to check whether the dual space approach is more useful 
than the ordinary perturbative approach.
The first fit function is calculated by the RG-improvement in the original $t$-space, which is given by\footnote{
In order to subtract the IR divergence in the same prescription as in the DSRS method,
we proceed as follows.
The IR divergence takes the form $\sim 1/(u+\delta)$ in $B_{C_2^{(h)}}$ of Eq.~\eqref{B_Cnh},
and the Borel integral of $1/(u+\delta)$ gives the $\log\delta$ singularity.
Hence, we subtract the contribution
from $B_{C_2^{(h)}}$ corresponding to $\log\delta$, expand in
$u$ and take the Borel integral.
We adjust the finite ($\delta$ independent) term such that the same divergence
is subtracted as in the DSRS method.
}
\begin{align}
I_{\rm fit}^{\rm RG}(t,s,\{k_i\};A,B)
=&\big[C^{(h)}_0(t)\big]_{\rm RG}^{(k_0)}+\big[C^{(h)}_1(t)\big]_{\rm RG}^{(k_1)}Am^2 t
\non &
+\Big[\big[C^{(h)}_2(t)\big]_{\rm RG}^{(k_2)}+2\log\delta\Big]A^2m^4t^2+Bm^4 t^2,
\label{fit-RG}
\end{align}
where
\be
\big[C^{(h)}_n(t)\big]_{\rm RG}^{(k_n)}=\big[C^{(h)}_n(t)\big]_{\rm PT}\bigg|_{{\rm up\, to\,}{\cal O}(\bar{g}(\mu^2)^{k_n+1})}\,{\rm with}~~~\mu^2=s^2/t
\,,
\ee
and $A={\langle{\alpha}\rangle}/{m^2}$ and 
$B={\langle\delta{\alpha}^2\rangle_{\rm PV}^{\rm ren}}/{m^4}$ stand for the fit parameters.
$[C_n^{(h)}]_{\rm PT}$ denotes the (naive)
perturbative expansion of $C_n^{(h)}$
in $\bar{g}(\mu^2)$.

The second one is calculated using the DSRS method and given by 
\begin{align}
I_{\rm fit}^{\rm PV}(t,s,\{k_n\};A,B)
=&
\big[C^{(h)}_0(t)\big]_{\rm PV}^{(k_0)}+\big[C^{(h)}_1(t)\big]_{\rm PV}^{(k_1)}Am^2 t
\non &
+\big[C^{(h)}_2(t)\big]_{\rm PV}^{(k_2)}A^2m^4t^2+Bm^4 t^2
\,,
\label{fit-PV}
\end{align}
where
we truncate the integrands of
Eqs.~\eqref{simint1}--\eqref{simint3} at ${\cal O}(\bar{g}(\mu^2)^{k_n+1})$ with
\be
\mu^2=s^2p^2.
\ee
(We treat $G^{\rm IR}$ exactly.)
In both formulas, we subtract the IR divergence of the ${\cal O}(t^2)$ term in the OPE, and each Wilson coefficient is calculated by truncation of the perturbative expansion.

In Eq.~\eqref{fit-RG}, the Wilson coefficients contain renormalons which cause uncertainties 
in the fit.
The major uncertainty comes from the $u=2$ renormalon in $\big[C^{(h)}_0(t)\big]_{\rm RG}$, which would give an order $100$ per cent 
uncertainty in the determination accuracy of $B$.
On the other hand, in Eq.~\eqref{fit-PV}, the Wilson coefficients are 
free of renormalons.
Hence, it is expected that we can extract
the exact values of the parameters,
\be 
r=1,~~~A=1,~~~B=-1.386\cdots
\,,
\label{exact-vals-param}
\ee
accurately by the fit.

\begin{figure}[tbp]
\centering
\includegraphics[width=15cm]{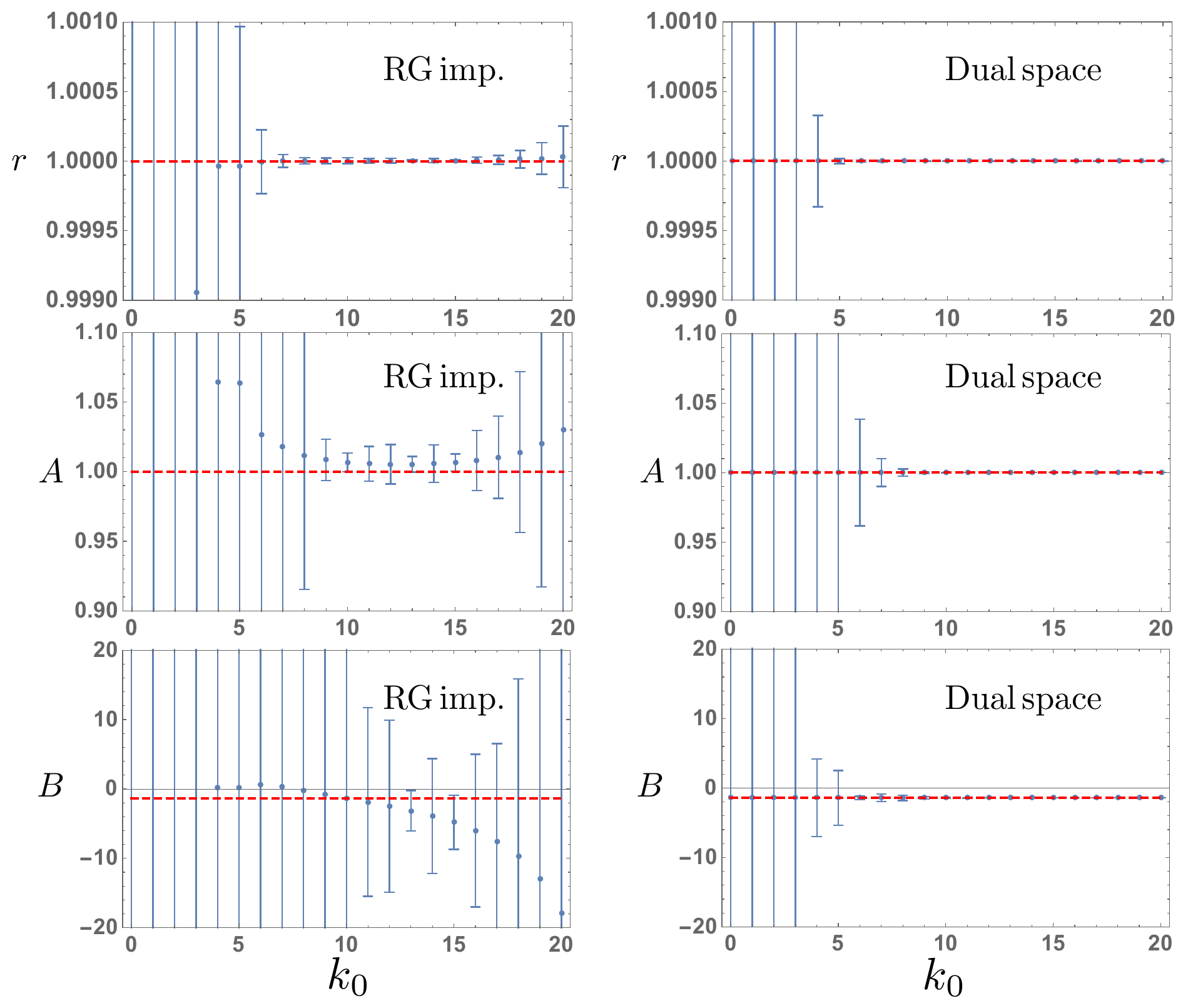}
\caption{\small
Results of the determination of the nonperturbative parameters $r$, $A$, $B$ by the fit.
The left (right) panels correspond to the Wilson coefficients $C^{(h)}_n$
calculated by the RG-improvement (DSRS method).
$k_0$ (horizontal axis) represents the truncation order  of the perturbative expansion
of $C^{(h)}_0$.
The blue dots with error bars represent the results of the fit by
varying the scale from the central value ($s=1$) by factors $1/2$ and $2$. 
The red dashed lines represent the exact values of the parameters.
}
\label{RG_dual}
\end{figure}

The details of the fits are as follows. 
The data points used for the fits correspond to 
$\hat{t}\equiv m^2t={n}/{10^4}$ for $n=1,\,2,\,\cdots,\,10$.
The maximal size of the coupling constant in this region is 
$\bar{g}={1}/{\log(10^3)}\approx0.14$.
The parameters $r,\,A$ and $B$ are determined simultaneously by the method of
least squares. 
For each Wilson coefficient, only $k_0$, the truncation order of $C^{(h)}_0$, is varied to examine the accuracy of the parameter determination. 
The systematic uncertainty is evaluated from the scale variation of $C^{(h)}_0$. 
The central value of the parameters is taken from the calculation with $s=1$ in 
Eqs.~\eqref{fit-RG} and \eqref{fit-PV}, and the systematic uncertainty 
is assigned from the difference 
as we vary $s\in\{1/2,1,2\}$. 
In this analysis, the truncation errors of perturbative expansions 
of the Wilson coefficients $C^{(h)}_n$ for $n\ge1$ are assumed to be negligible.
Hence, we fix the truncation orders of $C^{(h)}_1$ and $C^{(h)}_2$ at $k_1=3$ and $k_2=3$,
respectively,
and fix the scale at $s=1$ in both calculations, for simplicity.\footnote{
In the DSRS method, it is equivalent to fixing the truncation orders
at $k_1=1$ and $k_2=2$, since $d_{1,k}(0)=0$ for $k\ge 2$ and $d_{2,k}(0)=0$ for $k\ge 3$. 
}
In the analysis with Eq.~\eqref{fit-PV}, we assign another systematic uncertainty 
from (supposedly)
unknown ${\cal O}(t^3)$ nonperturbative corrections in the OPE. 
The uncertainty is estimated by the difference of the central values of the parameters 
$r,\,A$ and $B$ determined by the fits,
using $I_{\rm fit}^{\rm PV}$ and $I_{\rm fit}^{\rm PV}+Cm^6t^3$, in which $C$ is 
an additional fit parameter. 
This effect is negligible for the analysis with Eq.~\eqref{fit-RG} because 
the renormalon effect of order $m^4t^2$ is large.

The results of the fits are shown in Fig.~\ref{RG_dual}.
The blue dots with error bars represent the parameters determined
with different truncation orders $k_0$.
The red dashed lines show the exact values of the parameters.
In the RG-improvement method, the accuracy seemingly improves up to $k_0\sim13$ for all 
the parameters, but the uncertainty increases beyond that order. 
Since the $u=2$ renormalon is included in $C^{(h)}_0$, 
we expect that the perturbative series converges up to $k_0\approx k_0^*\equiv
2/\bar{g}\approx13.8$, which is consistent with the result.
While $r$ and $A$ are obtained with reasonable accuracy around $k_0=k_0^*$ even under the influence of the $u=2$ renormalon, 
$B$ is affected by an order $100$ per cent uncertainty in the determination. 
This feature is also consistent with the effects of the $u=2$ renormalon.
The determined values of $r,\,A$ and $B$ at $k_0=13\approx k_0^*$ are given by
\be
r-1=4(6)\times10^{-6},\quad A-1=5(6)\times10^{-3},\quad B=-3.1(2.9),
\label{res-RG}
\ee
which shows the limit of accuracy of the parameter
determination in the presence of the renormalon.
In general, if we use data points from a smaller (larger) $t$ region, we can determine
$r$ and $A$ with better (worse) accuracy, while $B$
is always subject to an order $100$ per cent uncertainty.

In the analysis using the DSRS method, 
the perturbative uncertainties decrease even beyond $k_0=k_0^*$. 
This is because all the renormalons are removed from $C^{(h)}_0$. 
Moreover, the convergence is 
even accelerated
before reaching the order $k_0^*$, where the original series is already dominated
by the LO renormalon contribution. 
On the other hand,
the total uncertainty never becomes smaller than a certain bound due to 
the systematic uncertainties by the higher-order of the OPE, namely the $t$ expansion.
The determined values of $r,\,A$ and $B$ at $k_0=13\approx k_0^*$ are given by
\be
r-1=4(4)\times10^{-8},\quad A-1=4(4)\times10^{-5},\quad B=-1.406(19),
\label{res-PV}
\ee
which are consistent with the exact values of the parameters,
Eq.~\eqref{exact-vals-param}.
We can see that the error sizes of $r$ and $A$ in Eq.~\eqref{res-PV} are orders of
magnitude
smaller than those in Eq.~\eqref{res-RG}.
This is because the major source of systematic uncertainties in Eq.~\eqref{res-PV} 
is the ${\cal O}(t^3)$ OPE corrections, while that in Eq.~\eqref{res-RG} originates from 
the $u=2$ renormalon in $\big[C^{(h)}_0\big]_{\rm RG}$ which is order ${\cal O}(t^2)$.
The results show that the DSRS method is useful to determine the nonperturbative 
parameters accurately by eliminating renormalons in the entire OPE. 
The theory prediction by the OPE can be made accurate
using the determined parameters.

\subsection{\boldmath Self-energy of $\sigma$}
\label{sec3:SEofSigma}

\begin{figure}[tbp]
\centering
\includegraphics[width=5cm]{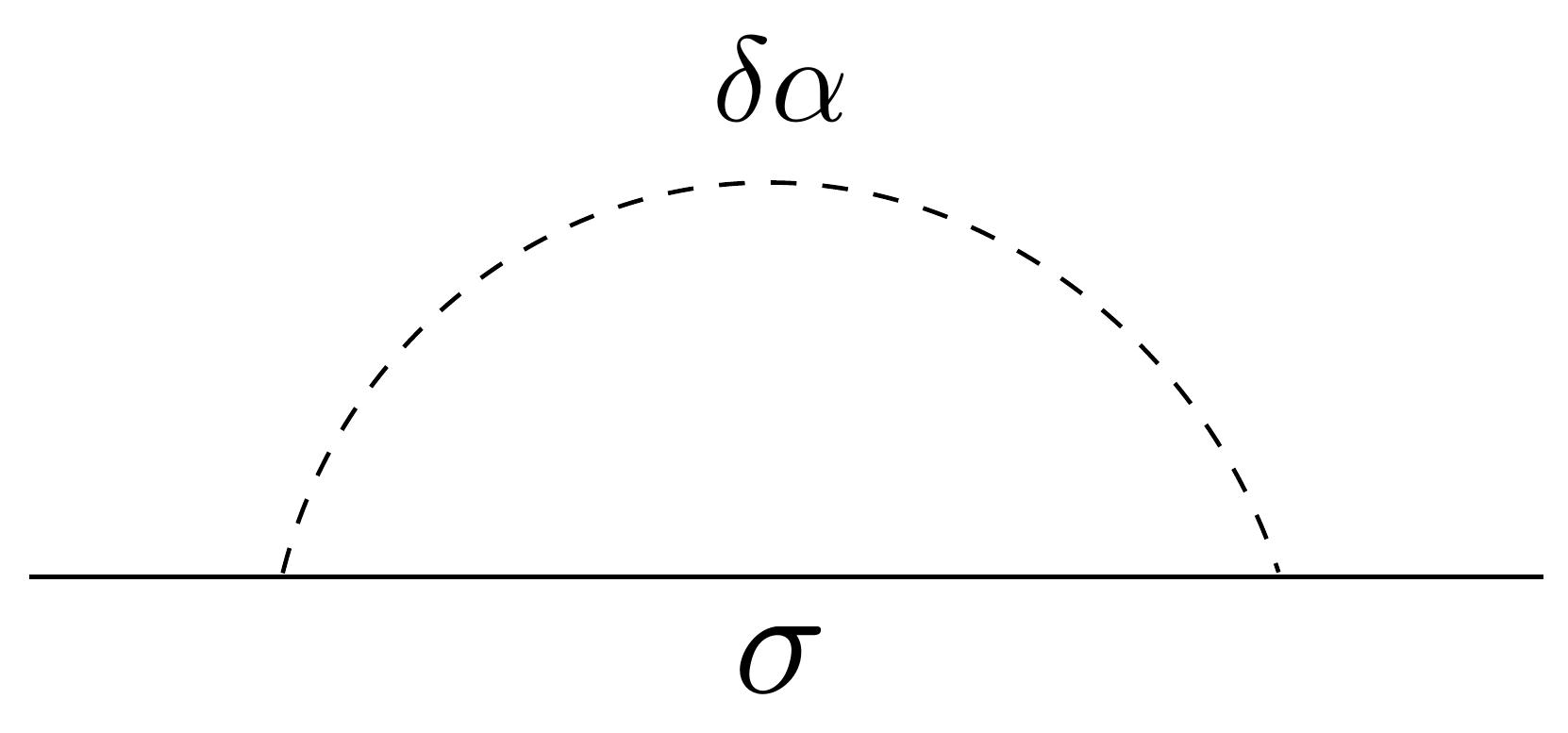}
\caption{\small
Self-energy diagram of $\sigma$ at
${\cal O}(1/N)$.
The tadpole diagram is omitted.
}
\label{Fig-SEgraphs}
\end{figure}

The inverse propagator of $\sigma$ with momentum $q^\mu$
is given by $(q^2+m^2)$ at 
LO of the $1/N$ expansion.
The second example we consider is the self-energy of $\sigma$ at
${\cal O}(1/N)$, denoted as $\Pi_\sigma(q^2)$.
The self-energy is included universally in calculations of 
correlation functions of $\sigma$ and is
given by the diagram in Fig.~\ref{Fig-SEgraphs}.
(We omit the tadpole diagram which is absorbed by
renormalization of $\langle\alpha\rangle$.)
Cancellation of the renormalons in the entire OPE of the self-energy
has been shown in ref.~\cite{Beneke:1998eq}.
Here, we show that the cancellation can be understood
systematically in the dual space approach.
The dual transform indeed eliminates all the IR renormalons.
We also see that the DSRS method is useful to improve
theoretical accuracy in the OPE.

In order to remove the UV divergences of $\Pi_\sigma(q^2)$,
we consider the second derivative of the self-energy,
\be
\rho(q^2)\equiv Nq^2\,\Pi_\sigma''(q^2)
=q^2\int \frac{d^2k}{(2\pi)^2} \left(\frac{\partial}{\partial q^2}\right)^2
\frac{1}{(q+k)^2+m^2}\, N\,D_\alpha(k^2)
\,.
\label{def-rho}
\ee
It is defined to be dimensionless and RG invariant.
The above integral is well defined nonperturbatively, without
UV or IR divergence.

We can construct the OPE of $\rho(q^2)$ in the dual space approach
similarly to the previous example.
We define the dual space observable $\tilde{\rho}(p^2)$ by
\be
\rho(q^2)=\left(\frac{1}{q^2}\right)^2
\int_{0}^\infty dp^2\,\tilde{\rho}(p^2)\, e^{-p^2/q^2}
\,.
\label{dual-transf-rho}
\ee
Using the generalized EBR, we can express the OPE of $\rho(q^2)$ 
as the sum of the hard 
and soft contributions:
\begin{align}
&
\big[\rho(q^2)\big]_{\rm OPE}
=\lim_{\delta\to 0}
\left\{ \rho^{\pm}_h(q^2)+\rho^{\pm}_s(q^2)\right\}
\,,
\label{EBR-rho(t)1}
\\
&
\rho^\pm_h(q^2)=\left(\frac{1}{q^2}\right)^{2-\delta}
\int_{C_\mp}dp^2\,
(p^2)^{-\delta_\pm}\left[\tilde{\rho}(p^2)\right]_{\rm OPE}\, e^{-p^2/q^2}
\,,
\label{EBR-rho(t)2}
\\
&
\rho^{\pm}_s(q^2)=\left(\frac{1}{q^2}\right)^{2-\delta}
\int_{C_\mp}dp^2\, (p^2)^{-\delta_\pm}
\tilde{\rho}(p^2)\,\left\{
1-\frac{p^2}{q^2}+\frac{1}{2}\left(\frac{p^2}{q^2}\right)^2-\dots
\right\}
\label{EBR-rho(t)3}
\,,
\end{align}
where the integrals are regularized in the same way as in the previous
example.

We can check that the Wilson coefficients of
$[\tilde{\rho}(p^2)]_{\rm OPE}$ 
do not have IR renormalons or IR divergences.
In fact, the Borel transform of
$[\tilde{\rho}(p^2)]_{\rm OPE}$ is given by
\begin{align}
B_{\tilde{\rho}}=&
-\frac{p^2}{\pi}\left(\frac{\mu^2}{p^2}\right)^u
\sum_{M=0}^\infty\left(\frac{m^2}{p^2}\right)^M
\Gamma(M+u-1)
\non&
\times
\sum_{n=0}^M \left(\frac{(n+u-1)_{M-n}}{(M-n)!}\right)^2
\,(-1)^{n}\,X_n(u)
\left\{ \sin(\pi u)\,\tilde{f}_{M,n}(p^2) + \cos(\pi u)\,\tilde{g}_{M}
\right\}
\,,
\label{BorelTr-tilde-rho}
\end{align}
where
\begin{align}
&
\tilde{f}_{M,n}(p^2) =
1-2(M\!+\!u)
\non &~~~~~~~~~~~~~
+(M\!+\!u)(M\!+\!u\!-\!1)\left\{
2\,\psi(M\!-\!n\!+\!1)-3\,\psi(M\!+\!u\!-\!1)+\log\!\left(\frac{p^2}{m^2}\right)\right\}
\,,
\\ &
\tilde{g}_{M}~~ =~~-2\pi (M\!+\!u)(M\!+\!u\!-\!1) 
\,,
\end{align}
and $\psi(z)=\Gamma^{\,\prime}(z)/\Gamma(z)$ denotes the digamma function.
(See App.~\ref{App-Calc-NLsigma} for the derivation.)
$B_{\tilde{\rho}}$ has no poles on the positive $u$ axis,
hence 
$[\tilde{\rho}(p^2)]_{\rm OPE}$ 
 has no IR renormalons.
(The poles at $u=0,1$ of the gamma and digamma functions for $M\leq 1$ are canceled
by other factors in the summand.)
On the other hand, $B_{\tilde{\rho}}$ has poles on the negative $u$ axis,
hence 
$[\tilde{\rho}(p^2)]_{\rm OPE}$ 
has UV renormalons.
The closest pole is located at $u=-1$.
It can destabilize convergence of the series expansion in the dual space.
Since the UV renormalons are Borel summable, we stabilize the
series by Borel resummation.
This is possible in the dual space approach, see App.~\ref{App-ResumUVrenormalon} for details.


$B_{\tilde{\rho}}$ has no pole at $u=0$ and is independent of any
regularization parameter.
Hence, $[\tilde{\rho}(p^2)]_{\rm OPE}$ is also free from IR divergences.
This aspect is the same as that of the previous example $\tilde{I}(p^2)$.

We can calculate the Borel transform of $\rho_h^\pm(q^2)$ by taking the
inverse dual transform of
$B_{\tilde{\rho}}$.
\begin{align}
B_{\rho_h}=&
-\frac{1}{\pi}\left(\frac{\mu^2}{q^2}\right)^u
\sum_{M=0}^\infty\left(-\frac{m^2}{q^2}\right)^M
\Gamma(M+u-1)
\non&
\times
\sum_{n=0}^M \frac{(n+u-1)_{M-n}}{\{(M-n)!\}^2}\,X_n(u)
\left\{ \sin(\pi u)\,{f}_{M,n}(q^2) + \cos(\pi u)\,{g}_{M}
\right\}
\non&~~~~~
~~~\times
\Gamma(2-n-u-\delta)
\,,
\label{BorelTr-rho_h}
\end{align}
where
\begin{align}
&
{f}_{M,n}(q^2) =
\tilde{f}_{M,n}(q^2) +
(M\!+\!u)(M\!+\!u\!-\!1)\,\psi(M\!+\!u\!-\!1)
\,,
\label{f(qsq)_Mn}
\\ &
{g}_{M}~~~ =~~~-\pi (M\!+\!u)(M\!+\!u\!-\!1) 
\,.
\end{align}
IR renormalons are generated by the inverse dual transform.
All of them stem from
the factor $\Gamma(2\!-\!n\!-\!u\!-\!\delta)$,
just as in the previous example, cf., Eq.~\eqref{B_Cnh}.
The generated IR renormalons have only integer powers
of $m^2/q^2$ and no logarithmic corrections are included.
This is because, the coefficient of $\log(q^2/m^2)$ term in Eq.~\eqref{f(qsq)_Mn}
has no poles of $u$ due to $\sin (\pi u)$ factor.
Hence, (a posteori) the simple dual transform Eq.~\eqref{dual-transf-rho}
is sufficient to remove all the renormalons.
The same factor $\Gamma(2\!-\!n\!-\!u\!-\!\delta)$ also generates IR divergences 
$\sim\log\delta$ for $M\ge 2$ by Borel integral.
The above formulas can be used to obtain the perturbative series
for the Wilson coefficients of $[\tilde{\rho}(p^2)]_{\rm OPE}$
and $[{\rho}(q^2)]_{\rm OPE}$.

Since $[\tilde{\rho}(p^2)]_{\rm OPE}$ does not have IR renormalons,
the dual space integral expression for $\rho^{\pm}_h(q^2)$
converges well to its exact value,
as we include more terms of the series expansion in $m^2/p^2$ and
$\bar{g}(p^2)$
(after resummation of the UV renormalons).
Using the dual space expressions for $\rho^{\pm}_h(q^2)$ and 
$\rho^{\pm}_s(q^2)$, 
the cancellation of the IR and UV divergences, as well as of the imaginary part,
can be shown in a similar manner to the previous example.
This is because, for instance, the UV behavior of the 
integrand of the soft contribution $\rho^{\pm}_s(q^2)$
is determined by $[\tilde{\rho}(p^2)]_{\rm OPE}$,
and hence, the general term necessary for the proof is of the form
$(p^2)^{n-\delta_\pm}\bar{g}(p^2)^k=(p^2)^{n-\delta_\pm}/\log(p^2/m^2)^k$
and the identity Eq.~\eqref{FundamentalId-EBR} can be used.
Furthermore, it is possible to perform a simulation study
and confirm good
convergence of the OPE to the exact $\rho(q^2)$
after renormalon subtraction by the DSRS method.
We perform a similar fit as in Sec.~\ref{sec3:simulation-study}
and obtain values of the nonperturbative parameters which
agree with the exact values with good accuracies.
Thus, apart from the resummation of the UV renormalons,
an analysis similar to the previous example can be made.

Finally we compare our study with the preceding
analyses on the OPEs in the $O(N)$ nonlinear $\sigma$ model.
Refs.~\cite{Bardeen:1976zh,David:1982qv,David:1985xj} 
studied renormalon cancellation in
the OPE of $\langle \delta \alpha^2 \rangle$
regularized by a cutoff, which mainly focused on
the LO Wilson coefficient and the ${\cal O}(m^4)$
matrix element.
Ref.~\cite{Beneke:1998eq} studied renormalon cancellation in the
OPE of the renormalized 
$\sigma$ self-energy.
It showed the renormalon cancellation as well as
cancellation between IR and UV divergences in the entire OPE in the Borel
space.
Ref.~\cite{Lee:2001ws} investigated an application 
of the cancellation of
the imaginary part calculated in ref.~\cite{Beneke:1998eq}
(up to ${\cal O}(m^2)$).
Novel aspects of the present paper are as follows.
First, we propose the new concept ``dual space approach,''
which provides a framework to analyze properties
of the OPEs including renormalons.
Secondly, using the framework, we show a method to
analyze cancellation of renormalons systematically
at arbitrary order of the expansion in the OPE
[using Eqs.~\eqref{FundamentalId-EBR} and \eqref{ren-Landau}],
for a general one-scale observable.
Thirdly, our method goes beyond 
diagrammatic analyses of renormalons,
since our analysis uses only the
dependence on the most external
single scale $p^2$,
without resolving detailed
kinematical information of individual diagrams.

\section{Application to QCD and discussion}
\label{sec5}

In this section we discuss how to subtract multiple
renormalons simultaneously from the OPE of an observable
in QCD.
We consider a procedure which can be used in practical
applications.
We also explain an unsolved question.

We first discuss the following points
in addition to the procedure used for the $O(N)$ nonlinear $\sigma$ model
in Sec.~\ref{sec3}:
(A) How to incorporate the effects of the fact that the form of each
renormalon can
deviate from a simple integer power of $\LQ$.
(B) How to incorporate the effects of the beta function beyond one loop order
($b_1,\,b_2,\,\cdots\ne 0$).

(A)
Renormalons are generally not in the simple power form of $\LQ/Q$ but
include corrections in the form of the anomalous dimension and
expansion in $\alfs(Q^2)$,
as given by  Eq.~\eqref{ren-form-fromOPE}.
A method for incorporating these corrections
is given in App.~B of ref.~\cite{Hayashi:2021vdq} in the context of the 
FTRS method.
Due to equivalence of the DSRS method to the
FTRS method combined with the resummation of the artificial UV renormalons,
we can also use that method in the dual space approach.

(B)
Including $b_1,b_2,\cdots\ne 0$ in the dual space
approach is straightforward,
and how this is done is already explained in Sec.~\ref{sec2} in principle.
Note that $\LMS$ in e.g.\ Eq.~\eqref{ren-form-fromOPE} is defined by the $k$-loop
beta function with $k$ not restricted to be one.
The points which may be worth notifying are as follows.
(i) In practical calculations, it is convenient to numerically obtain the 
running coupling constant in the complex $p^2$ plane by
solving the RG equation along the contour
$C_\pm$.
(ii)~IR divergences occur from the integral over $p^2$
from the IR region $p^2\sim 0$, beyond  a certain order in
expansion in $1/p$.
By  regularizing the $p^2$ integral
similarly to the nonlinear $\sigma$ model,
the IR divergences appear as $\sim\log\delta$ .
This is the case, even including NLL and beyond, since
the coupling constant behaves as
\be
\alfs(p^{2a})= \frac{1}{b_0\log(p^{2a}/\LMS^2)}-
\frac{b_1}{b_0^3}\frac{\log\log(p^{2a}/\LMS^2)}{\log(p^{2a}/\LMS^2)^2}
+{\cal O}\left(\log(p^{2a}/\LMS^2)^{-2}\right)
\,,
\ee
as $p^2\to 0$, and only the leading term causes the divergence.
For instance we can subtract the $\log\delta$ term as 
in Sec.~\ref{sec3}.

Let us add some more comments on the properties of the dual space
approach in QCD.

(1)
The conventional picture of renormalon cancellation in the
OPE is that the imaginary part of 
the (regularized) Wilson coefficients originating from the IR renormalons
are canceled by the imaginary part
contained in nonperturbative matrix elements of higher dimensional
operators.
In the dual space approach, 
the cancellation can be understood as follows.
The cancellation takes place at each order of the (simultaneous)
soft-{\it and}-hard expansion in the $p^2$ integral.
In the integrand, the soft expansion generates the expansion 
$e^{-tp^2}=1-tp^2+\cdots$, while
the hard expansion generates the OPE in the dual space,
$[\tilde{S}(p^2)]_{\rm OPE}$, which is the double expansion in $1/p^2$ and
$\alfs(p^{2a})$.
See Eq.~\eqref{ren-Landau}.
In the case $b_1,\,b_2,\,\cdots\ne 0$, the generalized EBR method
is guaranteed by the equality
\be
\int_{C_\pm} dp^2\, \frac{(p^2)^{a_\pm}}{\{\log(p^2/m^2)\}^n} \, [\log\log(p^2/m^2)]^k=0,
~~~~~(a_\pm \in \mathbb{C}, ~n,k \in \mathbb{Z}_{>0})
\label{FundamentalRel2-EBR}
\ee
since the high energy asymptotic expansion of $\alfs(p^{2a})$ generates such
a term.

(2)
In computing Wilson coefficients by perturbative QCD,
it is customary to use dimensional regularization.
Hence, instead of the pseudo-dimensional regularization
we used in Sec.~\ref{sec3} for simplicity, we can
use the full dimensional regularization.
This would help defining the nonperturbative matrix elements
universally, i.e., independently of observables, in a fully consistent
manner.

(3)
In the dual space approach,
we justify reconstruction of
an observable by the OPE
using the EBR relation Eq.~\eqref{FundamentalRel2-EBR}.
It means that the IR divergence should be treated in the $\log\delta$ (resummed)
form rather than the $1/\delta^n$ form which originates from the expansion of
$1/(u+\delta)$ in $u$ (in the Borel integral expression).
In particular only in the $\log\delta$ form can we show the cancellation 
of the imaginary part.
If we subtract IR divergences in the expanded form,
we are concerned about the ${\cal O}(\delta^0)$ (finite) term generated by 
the product of
$1/\delta^n$ and $\delta^n$.
Such terms remain and can be dependent on $Q^2$, 
which would give contributions different from the resummed prescription.
\medbreak

\begin{figure}[tbp]
\centering
\includegraphics[width=6cm]{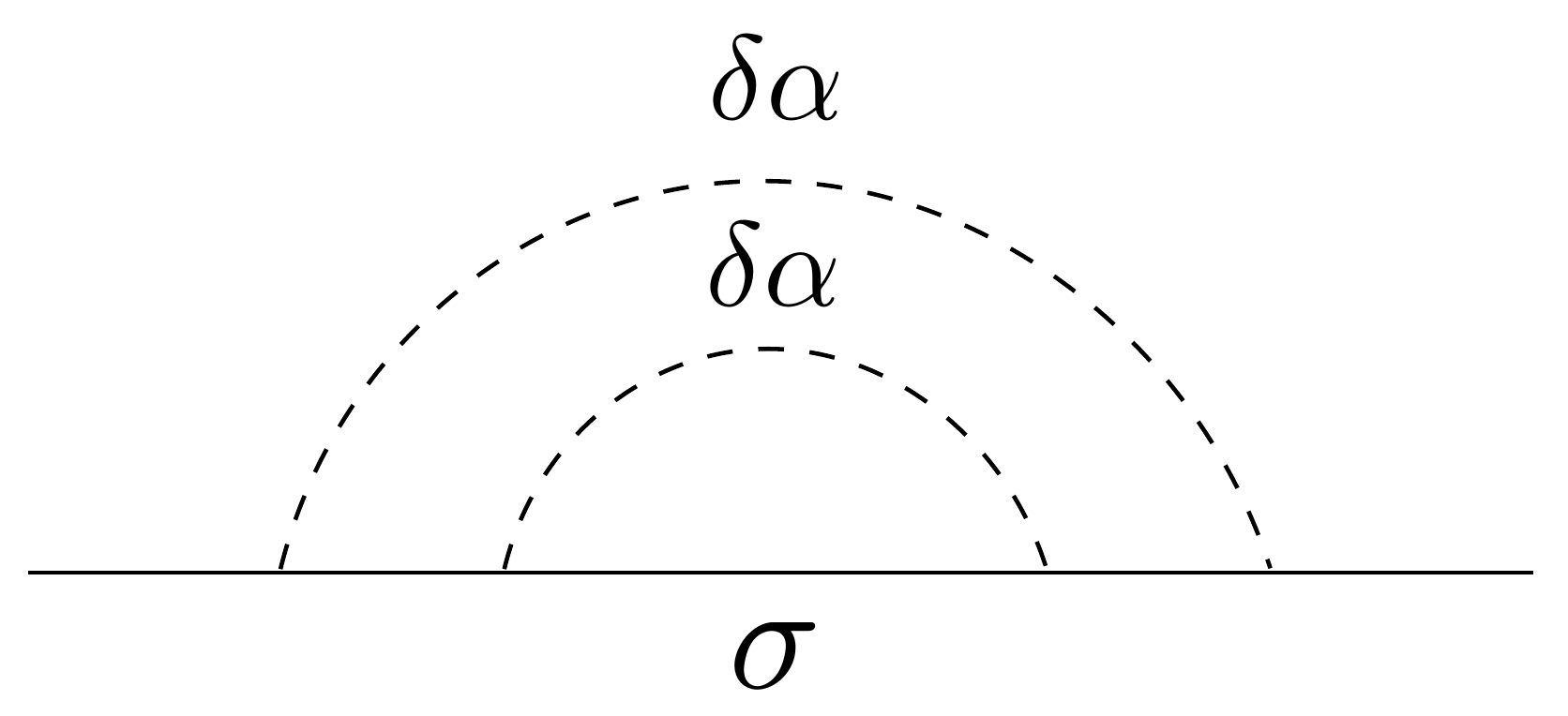}
\caption{\small
Example of a diagram which includes two $\delta\alpha$ propagators.
}
\label{Fig:Diagram-2Dalf}
\end{figure}
Finally
we discuss an unsolved problem.
In (A) above, we explained how to incorporate 
corrections to the simple power form $(\LQ/Q)^n$ of 
renormalons into the DSRS method.
We anticipate that the sum of
these corrections given as series
expansion in $\alfs(Q^2)$ possibly diverges.
For example, 
this type of corrections are expected to be included
in the diagram of Fig.~\ref{Fig:Diagram-2Dalf}
in the nonlinear $\sigma$ model, which contributes at ${\cal O}(1/N^2)$.
This diagram includes two $\delta\alpha$ propagators,
say, $D_\alpha(k_1^2)$ and $D_\alpha(k_2^2)$.
From the region where $k_1$ is hard and $k_2$ is soft,
$k_1\gg k_2\sim m$,
this diagram generates the product of the 
matrix element $\langle (\delta\alpha)^2 \rangle$, which
originates from $D_\alpha(k_2^2)$, and
Wilson coefficients $C_i$ corresponding to the contribution from
$[D_\alpha(k_1^2)]_{\rm OPE}$.
The Wilson coefficients $C_i$ include IR renormalons.
Thus, the LO Wilson coefficient, corresponding to
the (hard)-(hard) region of the diagram, is expected to include
a renormalon of the form $C_i\langle (\delta\alpha)^2 \rangle$,
where $C_i$ is given by a divergent series.
Although this problem is not specific to the
dual space approach,
up to now we do not know how to remove a renormalon
of this type using the DSRS method.
Stepwise inclusion of the corrections, first through
the anomalous dimension and then
order by order in $\alfs(Q^2)$ expansion,
is only an asymptotic approximation in this case.

\section{Summary and conclusions}

We proposed the DSRS method as a new method
to separate and subtract IR renormalons from 
Wilson coefficients in the OPE of a general one-scale observable $S(Q^2)$.
We constructed a dual transform from $S(Q^2)$ to $\tilde{S}(p^2)$
such that the OPE of $\tilde{S}(p^2)$ is free from renormalons.
The DSRS method has the following properties.
\begin{itemize}
\item[(a)] 
The DSRS method is equivalent to the conventional PV prescription
for subtracting IR renormalons,
which is based on the (regularized) Borel resummation technique.

\item[(b)]
The DSRS method is equivalent to the FTRS method after resumming
the artificially generated UV renormalons in the latter method.
In the DSRS method (unlike the FTRS method) no UV renormalons
are additionally generated by the dual transform.

\item[(c)]
The DSRS method can remove multiple IR renormalons simultaneously.

\item[(d)]
%

In the DSRS method
it is not required to evaluate the normalization constants of 
renormalons.
The real and imaginary part of the OPE are separately calculable
by the given procedure.

\item[(e)]
Wilson coefficients are calculated
by integrating series expansions in 
$\alfs(p^{2a})$ along the deformed contour
in the $p^2$ plane.
As a result, this method is insensitive
to instabilities due to the 
Landau singularity of the running
coupling constant.
(Note that,
with the current construction of
the dual transform, this property is limited to the case where
the renormalons are given by integer powers of
$\LQ/Q$.)

\item[(f)]

The DSRS method can give the PV result approximately 
with a given finite number of perturbative coefficients. 
The accuracy improves as more perturbative
coefficients are included.
This feature is of use in practical applications to QCD.


\item[(g)]
Using the DSRS method, we can determine nonperturbative
matrix elements accurately by a fit to experimental data
or lattice results.

\end{itemize}
The above properties of the DSRS method are derived
under the following assumptions.
We assumed that (i)  the OPE  of $S(Q^2)$ or $\tilde{S}(p^2)$ coincides with its
high-energy expansion, (ii)
for each Wilson coefficient, we can neglect
contributions from singularities other than renormalon singularities
in the right-half Borel plane, and
(iii) in the case of QCD, 
the beta function beyond five loops does not alter
the analytic structure of the beta function known up to five loops
\cite{Hayashi:2021vdq}.

In particular, the EBR method is applied
to the $p^2$ integral (inverse dual transform).
We emphasize that it constitutes the basis of
our analysis of the OPE.

\begin{itemize}
\item[(1)]
We applied the EBR method to the Laplace ($p^2$)
integral representation of $S(Q^2)$, which generates the
OPE of $S(Q^2)$.
The one-parameter integral form allows a simple
proof of the EBR method.

\item[(2)]
%
In the entire OPE,
IR and UV divergences cancel between the contributions from
the hard and soft regions.
The imaginary part cancel among the contributions from
the hard, soft and Landau singularity regions.
The cancellation can be understood
at each order of the (simultaneous)
soft-{\it and}-hard expansion, irrespective of the details of the observable.
\end{itemize}

We checked the above features by applying the DSRS method to some
observables in the 2D $O(N)$ nonlinear $\sigma$ model in the large $N$ limit.
In addition to their detailed examinations, 
we find that IR divergences in the Wilson coefficients
are removed in the dual
space, together with IR renormalons.
As a result, we can subtract the IR divergences in the original
space in an RG invariant manner
using the $p^2$ integral.

We discussed application of the DSRS method to QCD.
In principle, the method can be applied straightforwardly.
The equality that guarantees cancellation of renormalons
is given by Eq.~\eqref{FundamentalRel2-EBR}, which includes
subleading logarithmic corrections of the running coupling
constant.
We note that the FTRS method has already been applied to 
several QCD observables and good agreement with theoretical expectation
has been confirmed~\cite{Hayashi:2021vdq}.
We anticipate similar features for the DSRS method.

In conclusion, the DSRS method gives approximation of the OPE from
a finite number of perturbative expansions of Wilson coefficients,
with subtraction of multiple renormalons.
Hence, the method is expected to be simple
and efficient in phenomenological
applications, for improving accuracy of theoretical calculations. 

There remains an unsolved problem, which is not specific to the DSRS method,
when the correction to the power behavior $(\LQ/Q)^n$
of each IR renormalon is incorporated by anomalous dimension and
expansion in $\alfs(Q^2)$.
Such a correction can be an asymptotic series and we do not know
how to resum the series systematically.
This is a subdominant effect and at present it would be a less
serious question phenomenologically.
Nevertheless, it would be a subject that needs to be solved
toward realization of high precision theoretical calculations
in the future.

\section*{Acknowledgments}
Y.H.\ acknowledges support from GP-PU at Tohoku University. 
This work was supported by JSPS KAKENHI Grant Numbers
JP21J10226, JP20J00328, JP20K03923, JP19K14711, 
and JP18H05542.

\newpage

\newpage
\appendix
\section{Relation to FTRS method}
\label{app:RelToFTRS}

In this appendix we show that after resummation of artificial UV renormalons
in the FTRS method, it agrees with the DSRS method.

Let us consider a dimensionless and RG invariant observable $S(Q^2)$.
The starting point of the FTRS method is to define the Fourier transform of
$S(Q^2)$ as
\be
\tilde{S}'(\tau)=\int d^3\vec{x}\,e^{-i\vec{\tau}\cdot\vec{x}}r^{2au'}S(r^{-2a})
\,, 
~~~~~Q=r^{-a}\,,~~~|\vec{x}|=r\,,~~~|\vec{\tau}|=\tau
\,.
\ee
This is applied to the perturbation expansion $S(Q^2)=\sum_{n=0}^\infty c_n(L_Q)\alpha_s(\mu^2)^{n+1}$. 
Using
\be
S(r^{-2a})=\big(\mu^2r^{2a}\big)^{\hat{H}}\sum_{n=0}^\infty c_n(0)\alpha_s(\mu^2)^{n+1},
\ee
as the formal solution to the RG equation, 
with the operator $\hat{H}$ defined in Eq.~\eqref{opHinRGE},
we derive
\bea
\tilde{S}'(\tau)
&=&\int d^3\vec{x}\,e^{-i\vec{\tau}\cdot\vec{x}}r^{2au'}\big(\mu^2r^{2a}\big)^{\hat{H}}\sum_{n=0}^\infty c_n(0)\alpha_s(\mu^2)^{n+1}\non
&=&-\frac{4\pi}{\tau^{3+2au'}}\big(\mu^2/\tau^{2a}\big)^{\hat{H}}\sin(\pi a(\hat{H}+u'))\Gamma(2a(\hat{H}+u')+2)\sum_{n=0}^\infty c_n(0)\alpha_s(\mu^2)^{n+1}.
\nonumber\\
\label{AppA-FT1}
\eea
This expression shows that in the Fourier transformed series,  renormalons
at $u=-u'+\frac{n}{a}$ are suppressed
(one can see this by a similar computation to
Eqs.~\eqref{eq:suppIRrenom0} and \eqref{eq:suppIRrenom1}), while renormalons at $u=-u'-\frac{2n+1}{2a}$ 
are newly created,
where $n=0,\,1,\,2,\,\cdots$.
With a natural parameter choice, $a>0,\,u'<0$, IR renormalons are suppressed 
and UV renormalons are generated.
The latter renormalons are called ``artificial UV renormalons.''

First we note the identity to resum these aritificial UV renormalons:
\bea
\sin(\pi x)\Gamma(2x+2)
&=&-\frac{\pi\Gamma(2x+2)}{\Gamma(-x)\Gamma(1+x)}
=-\frac{\sqrt{\pi}}{\Gamma(-x)}2^{1+2x}\Gamma(x+3/2)\non
&=&-\frac{\sqrt{\pi}}{\Gamma(-x)}2^{1+2x}\int_0^\infty dy\,y^{x+1/2}\,e^{-y}
\,.
\eea
Using this identity we rewrite Eq.~\eqref{AppA-FT1} as
\bea
\tilde{S}'(\tau)
&=&\frac{4\pi^{3/2}}{\tau^{3+2au'}}\big(\mu^2/\tau^{2a}\big)^{\hat{H}}\frac{2^{1+2a(\hat{H}+u')}}{\Gamma(-a(\hat{H}+u'))}\int_0^\infty \!\!\!dy\,y^{a(\hat{H}+u')+1/2}\,e^{-y}\sum_{n=0}^\infty c_n(0)\alpha_s(\mu^2)^{n+1}\non
&=&\frac{4\pi^{3/2}}{\tau^{3+2au'}}\big(\mu^2/\tau^{2a}\big)^{\hat{H}}\frac{2^{1+2au'}}{\Gamma(-a(\hat{H}+u'))}\int_0^\infty \!\!\!dy\,y^{au'+1/2}\,e^{-y}\sum_{n=0}^\infty c_n(0)\alpha_s(\mu^2/(4y)^a)^{n+1}\non
&=&\frac{4\pi^{3/2}}{\tau^{3+2au'}}\int_0^\infty\!\!\! dy\,y^{au'+1/2}\,e^{-y}\frac{2^{1+2au'}}{\Gamma(-a(\hat{H}+u'))}\sum_{n=0}^\infty c_n(0)\alpha_s(\tau^{2a}/(4y)^a)^{n+1}
\,.
\eea
Here, we used the  RG invariance of $S$.

The regularized $S(Q^2)$ is obtained by the inverse Fourier transform
of $\tilde{S}'(\tau)$ with contour deformation.
Hence,
the principal value integral that eliminates renormalons are given by
\bea
\big[S(Q^2)\big]_{\rm PV}
&=&\frac{r^{-2au'-1}}{2\pi^2}\int_{0,\rm PV}^\infty d\tau\,\tau\sin(\tau r)\tilde{S}'(\tau)\non
&=&\frac{r^{-2au'-1}}{\sqrt{\pi}}\int_{0,\rm PV}^\infty \frac{d\tau}{\tau^{2+2au'}}\,\sin(\tau r)\int_0^\infty dy\,y^{au'+1/2}\,e^{-y}\frac{2^{2+2au'}}{\Gamma(-a(\hat{H}+u'))}
\non&&
~~~~~~~~~~~~~~~~~~~~~~~~~~~~
~~~~~~~~~~~~~~~~~~~~~~~~~~~~
\times
\sum_{n=0}^\infty c_n(0)\alpha_s(\tau^{2a}/(4y)^a)^{n+1}\non
&=&\frac{2r^{-2au'-1}}{\sqrt{\pi}}\int_0^\infty dy\,\,e^{-y}\int_{0,\rm PV}^\infty \frac{d\tau}{\tau^{2+2au'}}\,\frac{\sin(2\sqrt{y}\tau r)}{\Gamma(-a(\hat{H}+u'))}\sum_{n=0}^\infty c_n(0)\alpha_s(\tau^{2a})^{n+1}\non
&=&{2r^{-2au'}}\int_{0,\rm PV}^\infty \frac{d\tau}{\tau^{1+2au'}}\,e^{-\tau^2 r^2}\frac{1}{\Gamma(-a(\hat{H}+u'))}\sum_{n=0}^\infty c_n(0)\alpha_s(\tau^{2a})^{n+1}\non
&=&t^{-au'}\int_{0,{\rm PV}}^\infty dp^2\,e^{-tp^2}\frac{1}{(p^2)^{1+au'}}\frac{1}{\Gamma(-a(\hat{H}+u'))}\sum_{n=0}^\infty c_n(0)\alpha_s(p^{2a})^{n+1}
\,.
\eea
In the third equality the variable of integration is changed, and 
in the fourth equality we perform the $y$ integration.
Finally, we define $\tau=p,\,r^2=t$ for clarity.
This is just in the form of the Laplace transform from $t$-space to $p^2$-space and is an expression equivalent to the DSRS method.
See Eqs.~\eqref{tildeC0_PT} and \eqref{C0_PV+deltaC0}.

\section{Resummation of UV renormalons in DSRS method}
\label{App-ResumUVrenormalon}
In this appendix, we give a resummation formula for UV renormalons 
in the case that they are given by simple poles in the Borel plane
and in the LL approximation.
The formula is used in the analysis of Sec.~\ref{sec3:SEofSigma}.
In the formula, we introduce a new parameter $\bar{u}$ to control the resummation.
Let us take the LO Wilson coefficient in the dual space $\tilde{C}_0(p^2)$ and explain
how to resum renormalons.
$\tilde{C}_0(p^2)$ can be expressed by the following one-parameter integral form,
\bea
\tilde{C}_0(p^2)
&=&\frac{1}{(p^2)^{1+au'}}\sum_{n=0}^\infty \alpha_s(\mu^2)^{n+1}\tilde{c}_n(L_p)\non
&=&\frac{1}{(p^2)^{1+au'}}\frac{\Gamma(a(\bar{u}+\hat{H}))}{\Gamma(a(\bar{u}+\hat{H}))}\sum_{n=0}^\infty \alpha_s(\mu^2)^{n+1}\tilde{c}_n(L_p)\non
&=&\frac{1}{(p^2)^{1+au'}}\int_0^\infty d\zeta\,\zeta^{a(\bar{u}+\hat{H})-1}e^{-\zeta}\,\sum_{n=0}^\infty \alpha_s(\mu^2)^{n+1}\tilde{c}'_n(L_p)\non
&=&\frac{1}{(p^2)^{1+au'}}\int_0^\infty d\zeta\,\zeta^{a\bar{u}-1}e^{-\zeta}\,\sum_{n=0}^\infty \alpha_s(p^{2a}\zeta^{-a})^{n+1}\tilde{c}'_n(0)
\,.
\label{til-C0-UVresum}
\eea
Here, $\tilde{c}'_n$'s can be determined from $\tilde{c}_n$
or from the perturbative series in the original space, 
Eq.~\eqref{eq:c0pt1},
by
the following relation 
\begin{align}
\sum_{n=0}^\infty \alpha_s(\mu^2)^{n+1}\tilde{c}'_n(0)
&
=\frac{1}{\Gamma(a(\bar{u}+\hat{H}))}\sum_{n=0}^\infty \alpha_s(\mu^2)^{n+1}\tilde{c}_n(0)
\non &
=\frac{1}{\Gamma(a(\bar{u}+\hat{H}))\Gamma(-a(u'+\hat{H}))}\sum_{n=0}^\infty \alpha_s(\mu^2)^{n+1}c_n(0)
\,.
\end{align}
See Eqs.~\eqref{C0PT}--\eqref{til-C0-PT2}.
In the last line of Eq.~\eqref{til-C0-UVresum}, after applying $\zeta^{a\hat{H}}$ to $\alfs(\mu^2)^{n+1}$, we have set
$\mu^2=p^2$.
Note the RG running formula
\be
(\mu_1^2/\mu_2^2)^{\hat{H}}\alpha_s(\mu_1^2)^{n+1}=\alpha_s(\mu_2^2)^{n+1}
\,.
\label{App-resumUVren-running-formula}
\ee
It can be seen that the UV renormalons of $\sum_{n=0}^\infty \alpha_s(\mu^2)^{n+1}\tilde{c}_n(0)$ corresponding to 
$u=-\bar{u}$, $-\bar{u}-1/a$, $-\bar{u}-2/a$, $\cdots$ are suppressed by the $1/\Gamma(a(\bar{u}+\hat{H}))$ factor.

Then $\big[C_0\big]_\pm$ with resummation of UV renormalons is calculated by
\bea
\big[C_0(Q^2)\big]_\pm
&=&Q^{2u'}\int_{C_\mp} dp^2\,e^{-t p^2}\tilde{C}_0(p^2)\non
&=&Q^{2u'}\int_{C_\mp} dp^2\,e^{-t p^2}\frac{1}{(p^2)^{1+au'}}\int_0^\infty d\zeta\,\zeta^{a\bar{u}-1}e^{-\zeta}\,\sum_{n=0}^\infty \alpha_s(p^{2a}\zeta^{-a})^{n+1}\tilde{c}'_n(0)\non
&=&Q^{2u'}\int_0^\infty d\zeta\,\zeta^{a(\bar{u}-u')-1}e^{-\zeta}\int_{C_\mp} d\bar{p}^2\,e^{-\zeta t \bar{p}^2}\frac{1}{(\bar{p}^2)^{1+au'}}\,\sum_{n=0}^\infty \alpha_s(\bar{p}^{2a})^{n+1}\tilde{c}'_n(0)\non
&=&Q^{2u'}\int_{C_\mp} \frac{d\bar{p}^2}{(\bar{p}^2)^{1+au'}}\,\frac{\Gamma(a(\bar{u}-u'))}{(1+t \bar{p}^2)^{a(\bar{u}-u')}}\sum_{n=0}^\infty \alpha_s(\bar{p}^{2a})^{n+1}\tilde{c}'_n(0).
\label{C0pm-UVresum}
\eea
Recall that $t=Q^{-2/a}$.
In the third line we have changed the variable of integration from $p^2$ to
$\bar{p}^2=p^2\zeta^{-1}$.

%

\section{Proof of EBR method in Jantzen's manner}
\label{App-ProofEBR}

In this appendix we provide a proof of the EBR method
which we use in this paper.
We follow the strategy which was used by Jantzen in ref.~\cite{Jantzen:2011nz}.

Let us take the observable $I(t)$ studied in Sec.~\ref{sec3:gradient-flow} as an example:
\bea
I(t)&=&t^2\int_{0}^\infty dp^2\, \tilde{I}(p^2)\, e^{-tp^2}\non
&=&\lim_{\delta\to 0}~
t^{2-\delta}\int_{0}^\infty dp^2 \, p^{-2\delta_\pm}\, \tilde{I}(p^2)_\pm \, e^{-tp^2}
\,.
\label{App-Jantzen-1}
\eea
For convenience of notation, instead of deforming the contour to $C_\pm$,
we have taken the integral path on the
positive real axis and instead shifted the Landau pole infinitesimally
above or below the axis, $m^2_\pm\equiv m^2\pm i0$, and write
$\tilde{I}(p^2)_\pm \equiv\tilde{I}(p^2)|_{m^2\to m^2_\pm}$.
We note that $\tilde{I}(p^2)$ is regular at ${\rm Re}\,p^2>0$.

We introduce the factorization scale $\mu_F$
such that $\LQ^2\ll \mu_F^2 \ll t^{-1}$
and separate the integral region to above and below $\mu_F^2$:
\bea
I_{\rm reg}&\equiv&
t^{2-\delta}\int_{0}^\infty dp^2 \, p^{-2\delta_\pm}\, \tilde{I}(p^2)_\pm \, e^{-tp^2}
\non
&=&t^{2-\delta}\bigg(\int_{0}^{\mu_F^2}+\int_{\mu_F^2}^\infty\bigg) dp^2\, p^{-2\delta_\pm}\tilde{I}(p^2)_\pm \,e^{-tp^2}\non
&=&t^{2-\delta}\int_{0}^{\mu_F^2}dp^2 p^{-2\delta_\pm}\tilde{I}(p^2)_\pm T_{tp^2}\big[e^{-tp^2}\big]+t^{2-\delta}\int_{\mu_F^2}^\infty dp^2 p^{-2\delta_\pm}\big[\tilde{I}(p^2)_\pm\big]_{\rm OPE} e^{-tp^2}
\,,
\non
\eea
where $T_x[f(x)]$ represents the Taylor expansion of
$f(x)$ in $x$.
In the last line,
we expand the integrand using hierarchy of the variables
satisfied in each region.
($[\tilde{I}(p^2)_\pm]_{\rm OPE}$ represents double expansion
in $1/p^2$ and $1/\log(p^2/m_\pm^2)$.)

We further rewrite the integral regions:
\bea
I_{\rm reg}
&=&t^{2-\delta}\bigg(\int_{0}^{\infty}-\int_{\mu_F^2}^\infty\bigg)dp^2 \, p^{-2\delta_\pm}
\,\tilde{I}(p^2)_\pm \, T_{tp^2}\big[e^{-tp^2}\big]\non
&&\quad\quad+t^{2-\delta}\bigg(\int_{0}^{\infty}-\int_0^{\mu_F^2}\bigg) dp^2\,
 p^{-2\delta_\pm}\,\big[\tilde{I}(p^2)_\pm\big]_{\rm OPE}\, e^{-tp^2}\non
&=&t^{2-\delta}\int_{0}^{\infty}dp^2 p^{-2\delta_\pm}\tilde{I}(p^2)_\pm T_{tp^2}\big[e^{-tp^2}\big]+t^{2-\delta}\int_{0}^{\infty}dp^2 p^{-2\delta_\pm}\big[\tilde{I}(p^2)_\pm\big]_{\rm OPE} e^{-tp^2}\non
&&\quad\quad-t^{2-\delta}\bigg(\int_{0}^{\mu_F^2}+\int_{\mu_F^2}^\infty\bigg) dp^2 p^{-2\delta_\pm}\big[\tilde{I}(p^2)_\pm\big]_{\rm OPE} T_{tp^2}\big[e^{-tp^2}\big].
\eea
In the last expression, the first two terms are equal to $I_s^\pm(t)$
and $I_h^\pm(t)$, respectively.
The last term is the sum of terms in the form of Eq.~\eqref{FundamentalId-EBR}
or \eqref{FundamentalRel2-EBR}, 
which are zero by analytic continuation.
This completes the proof.

\section{\boldmath Calculations for OPEs in nonlinear $\sigma$ model}
\label{App-Calc-NLsigma}
In this appendix we present details of the calculation for the OPEs
of the observables in the nonlinear $\sigma$ model.

\subsection{Factorization formula for Borel transform}
First we present a formula which is useful in calculating Borel transform of a
perturbative series in the LL approximation, namely in the case that the one-loop
running is exact, such as in the nonlinear $\sigma$ model.

Consider an observable given as an integral form
\be
S(q^2)=\int \prod_i dx_i \int d^Dp \,\, W(x_i;q,p) \, G(p^2)
\,,
\label{App-calc-defS}
\ee
where $q,p$ are vectors in $D$ dimension.
$G(p^2)$ is given as a perturbative series
\be
G(p^2)=(p^2)^{-\nu} \sum_k  c_k(L_p)\, \bar{g}(\mu^2)^{k+1}
\,,
~~~~~
L_p=\log \left(\frac{\mu^2}{p^2}\right)
\,.
\ee
We assume both $S(q^2)$ and $G(p^2)$ are RG invariant.

We define an integral where $G(p^2)$ is replaced by a simple power
of $p^2$:
\be
F(a)\equiv
\int \prod_i dx_i \int d^Dp \,\, W(x_i;q,p) \, \left(\frac{\mu^2}{p^2}\right)^a
\,.
\label{App-calc-defF}
\ee
Then, the Borel transform of $S(q^2)$ is given in a factorized form as
\be
B_S(u)=F(\nu+u)\,B_G(u;p^2=\mu^2)
\,.
\label{App-BorelFacFormula}
\ee
This formula is useful in the case that one can calculate $B_G(u)$ and $F(a)$,
while 
direct evaluation of $B_S(u)$ via
Eq.~\eqref{App-calc-defS} is difficult.
\\
(Proof)
\\
By setting $\mu^2=p^2$ and
using the running formula
Eq.~\eqref{App-resumUVren-running-formula}, we can write
\begin{align}
&G(p^2)=(p^2)^{-\nu} \sum_k  c_k(0)\, \bar{g}(p^2)^{k+1}
=\left(\frac{\mu^2}{p^2}\right)^{\hat{H}+\nu}\sum_k  c_k(0)\, \bar{g}(\mu^2)^{k+1}
(\mu^2)^{-\nu} 
\,,
\\&
\hat{H}=\bar{g}^2\frac{\partial}{\partial \bar{g}}
\,, ~~~~~
\text{in particular,}~~~
\hat{H}\,\bar{g}^{k+1}=(k+1)\,\bar{g}^{k+2}
\,.
\end{align}
Inserting this into Eq.~\eqref{App-calc-defS} and using 
Eq.~\eqref{App-calc-defF}, we obtain
\begin{align}
S(q^2)&=F(\hat{H}+\nu)\sum_k  c_k(0)\, \bar{g}(\mu^2)^{k+1}\,(\mu^2)^{-\nu}
\non
&=\sum_j \frac{F^{(j)}(\nu)}{j!}\,\hat{H}^j
\sum_k  c_k(0)\, \bar{g}(\mu^2)^{k+1}\,(\mu^2)^{-\nu}
\non
&=\sum_j \frac{F^{(j)}(\nu)}{j!}\,
\sum_k  c_k(0)\, \frac{(k+j)!}{k!}\,
\bar{g}(\mu^2)^{k+j+1}\,(\mu^2)^{-\nu}
\,.
\label{App-calculate-SinH}
\end{align}
We take the Borel transform and obtain
\begin{align}
B_S(u)
&=\sum_j \frac{F^{(j)}(\nu)}{j!}\,u^j
\sum_k  \, \frac{c_k(0)}{k!}\,u^k
\,(\mu^2)^{-\nu}
\non&
=F(\nu+u)\,B_G(u;p^2=\mu^2)
\,.
\end{align}

\subsection{\boldmath Calculations of $B_{\tilde{I}}$, $X_n$, $B_{I_h}$}

We can calculate the Borel transforms $B_{\tilde{I}}$ and $X_n$ as follows.
For the moment we fix $\mu^2=p^2$.
Define
\bea
f(x,\bar{g})&\equiv&\frac{\tilde{I}(p^2)}{p^2}=\frac{\sqrt{1+4x}}{\log\Big(\frac{\sqrt{1+4x}+1}{\sqrt{1+4x}-1}\Big)}
=\sum_{n=0}^\infty x^n\sum_{k=0}^\infty d_{n,k}(0)\,\bar{g}(p^2)^{k+1}\,,
\eea
where
\bea
&& x=\frac{m^2}{p^2},~~~\,\bar{g}(p^2)=\frac{1}{\log(p^2/m^2)}
\,.
\eea
The Taylor expansion of $f(x,\bar{g})$ in $\bar{g}(p^2)$ is given by
\bea
f(x,\bar{g})&=&\frac{\sqrt{1+4x}}{\log\Big(\frac{(\sqrt{1+4x}+1)^2}{4x}\Big)}
=\frac{\bar{g}\sqrt{1+4x}}{1+2\bar{g}\log\Big(\frac{\sqrt{1+4x}+1}{2}\Big)}\non
&=&\sqrt{1+4x}\sum_{k=0}^\infty\bigg[-2\log\Big(\frac{\sqrt{1+4x}+1}{2}\Big)\bigg]^k\,\bar{g}^{k+1}.
\eea
The Borel transform of $f$ is readily obtained as
\be
B_f(u,x)=\sqrt{1+4x}\sum_{k=0}^\infty\frac{1}{k!}\bigg[-2\log\Big(\frac{\sqrt{1+4x}+1}{2}\Big)\bigg]^k\,u^{k}=\sqrt{1+4x}\bigg(\frac{\sqrt{1+4x}+1}{2}\bigg)^{-2u}
\,,
\ee
and $B_{\tilde{I}}=p^2 B_f$.

$X_n$ is included in $B_f$ as $B_f(u,x)=\sum_{n=0}^\infty X_n(u)\,x^n$.
Taylor expansion of $B_f$ in $x$ is calculated as follows:
\bea
B_f(u,x)&=&\sum_{m=0}^\infty\frac{4^u\Gamma(1-2u)}{\Gamma(m+1)\Gamma(1-m-2u)}(1+4x)^{\frac{m+1}{2}}\non
&=&\sum_{n=0}^\infty\sum_{m=0}^\infty\frac{4^{u+n}\Gamma(1-2u)\Gamma(3/2+m/2)}{\Gamma(m+1)\Gamma(1-m-2u)\Gamma(3/2+m/2-n)\Gamma(n+1)}x^n\non
&=&\sum_{n=0}^\infty\bigg[4^{n+u}\frac{\Gamma(3/2)\,_3F_2\big(\{3/2,\,u,\,1/2+u\},\{1/2,\,3/2-n\},1\big)}{\Gamma(3/2-n)\Gamma(1+n)}\non
&&\quad\quad+4^{n+u}\frac{\Gamma(1-2u)\,_3F_2\big(\{1+n,\,-1/2+n+u,\,n+u\},\{n,\,1/2+n\},1\big)}{\Gamma(2n)\Gamma(2-2n-2u)}\bigg]x^n\non
&=&\sum_{n=0}^\infty\bigg[\frac{\Gamma(-n-2u)}{\Gamma(n+1)\Gamma(2-2n-2u)}(4u^2-2u-2n)\bigg]x^n,
\eea
where we used an identity
\be
_3F_2(\{a+1,\,b,\,c\},\{a,\,d\},1)=\frac{\Gamma(d)\Gamma(d-b-c-1)}{\Gamma(d-b)\Gamma(d-c)}\Big(d-b-c-1+\frac{bc}{a}\Big).
\ee

Let us now calculate the Borel transform
$B_f(u,x)$ with respect to $\bar{g}(\mu^2)$, namely
in the case $\mu$ is left as an independent parameter.
In this case, using the running formula
Eq.~\eqref{App-resumUVren-running-formula}, we have
\bea
f(x,\bar{g})
=\sum_{n=0}^\infty x^n\sum_{k=0}^\infty d_{n,k}(0)\,\bar{g}(p^2)^{k+1}
=
\left(\frac{\mu^2}{p^2}\right)^{\hat{H}}
\sum_{n=0}^\infty x^n\sum_{k=0}^\infty d_{n,k}(0)\,\bar{g}(\mu^2)^{k+1}
\,.
\eea
This has the same form as the first line of Eq.~\eqref{App-calculate-SinH}
with $\nu=0$ and $F(a)=(\mu^2/p^2)^a$.
Hence,
\be
B_f(u,x;p^2;\mu^2)
=
\biggl(\frac{\mu^2}{p^2}\biggr)^{\! u}\,
B_f(u,x;\mu^2;\mu^2)
=\sqrt{1+4x}\bigg(\frac{\sqrt{1+4x}+1}{2}\bigg)^{\! -2u}
\biggl(\frac{\mu^2}{p^2}\biggr)^{\! u}
\,.
\ee

To calculate $B_{I_h}(u)$, we can apply the factorization formula
Eq.~\eqref{App-BorelFacFormula}
in the case where $\nu=n-1$,
$G(p^2)=p^2(m^2/p^2)^n\tilde{C}_n(p^2)$ [cf., Eqs.~\eqref{tildeI_OPE}, 
\eqref{WilsonCoeff-tildeI}], and
\be
F(a)=
t^{2-\delta}\!\int_{C_\mp}
dp^2\, (p^2)^{-\delta_\pm}
\,e^{-tp^2}
\left(\frac{\mu^2}{p^2}\right)^{a}
=(\mu^{2}t)^a\,t\,\Gamma(1-\delta-a)
\,.
\ee
Then, we readily obtain Eq.~\eqref{B_Cnh}.

\subsection{\boldmath Calculations of $B_{\rho_h}$, $B_{\tilde{\rho}}$}

There are three regions when the EBR method is applied to 
$\rho(q^2)$ defined by Eq.~\eqref{def-rho}: 
(i) $k\sim q \gg m$, (ii) $(q+k)\sim m \ll q,k$, and
(iii) $k \sim m \ll q$.
In order to regularize each integral in the expansion,
we employ dimensional regularization except for $D_\alpha(k^2)$
whose expression for $D=2$, Eq.~\eqref{Dalpha}, is used:\footnote{
This treatment is for the sake of simplicity, since 
keeping $D_\alpha(k^2)$ also in dimensional regularization makes
the calculations tedious.
In the end, the dual space quantities $\tilde{\rho}$ and
$B_{\tilde{\rho}}$ are obtained to be finite as $D\to 2$,
so this treatment should be justified.
}
\be
\rho(q^2)_{\rm reg}
\equiv q^2\left(\frac{\partial}{\partial q^2}\right)^2
\int \frac{d^Dk}{(2\pi)^D} 
\frac{1}{(q+k)^2+m^2}\, N\left[D_\alpha(k^2)\right]_{D=2}
\,.
\label{App-calc-def-rho}
\ee
To compute the dual space observable 
$[\tilde{\rho}(p^2)]_{\rm OPE}$, the region (iii) does not contribute.
This is because, it gives a series expansion with
only integer powers of $q^2$
(without $\log q^2$ terms), whose dual
transformation vanishes.

The sum of the contributions from the regions (i) and (ii) can be calculated
together as follows.
\bea
\rho(q^2)_{\rm reg}^{\rm (i,ii)}
&=& q^2\left(\frac{\partial}{\partial q^2}\right)^2\int \frac{d^Dk}{(2\pi)^D} 
\frac{1}{(q+k)^2+m^2}\, N\left[D_\alpha(k^2)\right]_{\rm OPE}
\non
&=& 4\pi \mu^2 \sum_{n,k}F(n+\hat{H}-1)
\left(\frac{m^2}{\mu^2}\right)^{n}
d_{n,k}(0)\,\bar{g}(\mu^2)^{k+1}
\,,
\eea
where
\begin{align}
F(a)&=
q^2\left(\frac{\partial}{\partial q^2}\right)^2
\int \frac{d^Dk}{(2\pi)^D} 
\frac{1}{(q+k)^2+m^2}\, 
\left(\frac{\mu^2}{k^2}\right)^{a}
\non
\rule{0mm}{10mm}
&=\frac{m^{-2\delta}}{(4\pi)^{1-\delta}}
\left(\frac{\mu^2}{m^2}\right)^{\! a}\,
\frac{\Gamma(1-a-\delta)\Gamma(a+\delta)}{\Gamma(1-\delta)}
\non&~~~~~~~~~~~~~~~~~
\times
q^2\left(\frac{\partial}{\partial q^2}\right)^2\,
_2F_1\left(a,a+\delta;1-\delta;-\frac{q^2}{m^2}\right)
\,.
\end{align}
We set $D=2-2\delta$.

Hence, using the factorization formula, we obtain the Borel transform
of $\rho_h^\pm(q^2)$ as
\be
B_{\rho_h}(u)=4\pi 
\mu^2
\sum_{n=0}^\infty F(n+u-1) \, X_n(u)\, 
\left(\frac{m^2}{\mu^2}\right)^{n}
\,.
\ee
We can rewrite $_2F_1(a,b;c;-\frac{q^2}{m^2})$
in terms of $_2F_1(a',b';c';-\frac{m^2}{q^2})$ using a functional
identity, by which one can readily find the all order formula of $F(n+u-1)$
in expansion in $m^2/q^2$.
All the IR renormalons and IR divergences of $\rho_h^\pm(q^2)$ originate from
the factor $\Gamma(2-n-u-\delta)$ in $F(n+u-1)$.
We can take the limit $\delta \to 0$ for the other part, which gives
the formula
Eq.~\eqref{BorelTr-rho_h}.

We can take the dual transform of Eq.~\eqref{BorelTr-rho_h} using 
Eq.~\eqref{EBR-rho(t)2}, order by order in expansion in $1/q^2$.
The dual transform essentially 
generates $1/\Gamma(2\!-\!M\!-\!u\!-\!\delta)$.
Since $\Gamma(2\!-\!n\!-\!u\!-\!\delta)/\Gamma(2\!-\!M\!-\!u\!-\!\delta)$ for $n\leq M$
is a polynomial
of $u$, all the singularities corresponding to the IR renormalons
and IR divergences are canceled
by the dual transform.
After the cancellation, we can safely take the limit $\delta\to 0$.
Thus, we
obtain $B_{\tilde{\rho}}$ given by Eq.~\eqref{BorelTr-tilde-rho}.
It is straightforward to obtain $[\tilde{\rho}(p^2)]_{\rm OPE}$ from
$B_{\tilde{\rho}}$ at each order of the expansion in $1/p^2$ and $\bar{g}(p^2)$.
We expect that the obtained result for $B_{\tilde{\rho}}(u)$ or
$[\tilde{\rho}(p^2)]_{\rm OPE}$ is independent of the
regularization scheme we employ at intermediate stage
of the calculation, due to the absence of IR renormalons and IR
divergences.\footnote{
The algorithm for deriving $[\tilde{\rho}(p^2)]_{\rm OPE}$
is partly cut short, for simplicity.
The true algorithm is as follows.
We first determine the regularized dual transform
which cancels all the IR renormalons and IR divergences
of $\rho_h^\pm(q^2)$, and afterwards
we remove the regularization.
Once $[\tilde{\rho}(p^2)]_{\rm OPE}$ is determined,
$\rho_h^\pm(q^2)$ can be redetermined regularization dependently.
The regularization dependence cancels against that of $\rho_s^\pm(q^2)$.
}

\end{document}